\documentclass[apj]{emulateapj}
\pdfoutput=1
\usepackage{hyperref}
\usepackage{amsmath,amstext}
\usepackage[T1]{fontenc}
\usepackage[figure,figure*]{hypcap}
\usepackage{tikz}
\usepackage[all]{nowidow}
\usetikzlibrary{shapes,arrows}

\shorttitle{A New Cool WD Atmosphere Code}
\shortauthors{Blouin, Dufour \& Allard}

\begin{document}

\submitted{Accepted for publication in The Astrophysical Journal}

\title{A New Generation of Cool White Dwarf Atmosphere Models.
  I. Theoretical Framework and Applications to DZ Stars}

\author{S. Blouin\altaffilmark{1}}
\author{P. Dufour\altaffilmark{1}}
\author{N.F. Allard\altaffilmark{2,3}}

\altaffiltext{1}{D\'epartement de Physique, Universit\'e de Montr\'eal, Montr\'eal,
  QC H3C 3J7, Canada; sblouin@astro.umontreal.ca, dufourpa@astro.umontreal.ca.}
\altaffiltext{2}{GEPI, Observatoire de Paris, Universit\'e PSL, CNRS, UMR 8111,
  61 avenue de l'Observatoire, 75014 Paris, France.}
\altaffiltext{3}{Sorbonne Universit\'e, CNRS, UMR 7095,
  Institut d'Astrophysique de Paris, 98bis boulevard Arago, 75014 Paris, France.}

\begin{abstract}
  The photospheres of the coolest helium-atmosphere white dwarfs are
  characterized by fluid-like densities. Under those conditions,
  standard approximations used in model atmosphere codes are no longer
  appropriate. Unfortunately, the
  majority of cool He-rich white dwarfs show no spectral features,
  giving us no opportunities to put more elaborate models to the
  test. In the few cases where spectral features are observed (such as
  in cool DQ or DZ stars), current models completely fail to reproduce
  the spectroscopic data, signaling shortcomings in our theoretical
  framework. In order to fully trust parameters derived solely from
  the energy distribution, it is thus important to at least succeed in
  reproducing the spectra of the few coolest stars exhibiting spectral
  features, especially since such stars possess even less extreme
  physical conditions due to the presence of heavy elements. In this
  paper, we revise every building block of our model atmosphere code
  in order to eliminate low-density approximations. Our updated white
  dwarf atmosphere code incorporates state-of-the-art
  constitutive physics suitable for the conditions found in cool
  helium-rich stars (DC and DZ white dwarfs). This includes new
  high-density metal line profiles, nonideal continuum opacities, an
  accurate equation of state and a detailed description of the
  ionization equilibrium. In particular, we present new ab initio
  calculations to assess the ionization equilibrium of heavy elements
  (C, Ca, Fe, Mg and Na) in a dense helium medium and show how our improved
  models allow us to achieve better spectral fits for two cool DZ
  stars, Ross 640 and LP 658-2.
\end{abstract}

\section{Introduction}
Pure helium-rich white dwarfs do not show any spectral lines when
$T_{\mathrm{eff}} \lesssim 10000\,$K. The same occurs for
$T_{\mathrm{eff}} \lesssim 5000\,$K in the case of pure hydrogen-rich
atmospheres. Together, these featureless white dwarfs are known as DC
stars. One is thus forced to rely solely on the shape of the spectral
energy distribution to deduce the chemical composition and effective
temperature of these white dwarfs
\citep{bergeron1997chemical,bergeron2001photometric}. Although most
cool white dwarfs have featureless spectra, some cool helium-rich
white dwarfs do show significant spectral features that can be
exploited to retrieve additional information on the physical
conditions encountered in their atmospheres. Some contain enough
hydrogen to show strong H$_2$-He collision-induced absorption (CIA)
features, some show C$_2$ Swan bands (DQ and DQpec stars) and others
show metal lines (DZ stars). Interestingly, in all cases, models fail
to reproduce these spectra. For instance, the CIA is inadequately modeled
\citep[e.g., \mbox{LHS 3250}, \mbox{SDSS J123812.85+350249.1},
  \mbox{SDSS J125106.11+440303.0},][]{gianninas2015ultracool}, the
C$_2$ bands are distorted \citep[e.g., LHS 290,][]{kowalski2010origin}
and the metal absorption lines often do not have the right strength or
the right shape (e.g., WD 2356-209, \citealt{bergeron2005interpretation,
homeier2005,homeier2007}, LP 658-2, \citealt{dufour2007spectral,wolff2002element}).

For all these stars, the discrepancies between models and observations
can be related to nonideal high-density effects arising at the
photosphere since for cool ($T_{\mathrm{eff}} < 6000\,$K) helium-rich
white dwarfs, densities reach fluid-like values. At a Rosseland
optical depth $\tau_R = 2/3$, density can be as high as
$1\,\mathrm{g\,cm}^{-3}$
\citep{bergeron1995new,kowalski2010understanding}, which corresponds
to a fluid where the separation between atoms is roughly equivalent to
the dimension of atoms themselves. Clearly, under such conditions,
interactions between species are no longer negligible and the ideal
gas approximation must be discarded.

The nonideal effects arising from this high density have remained
mostly unnoticed for DC stars, since a featureless spectrum provides
little opportunity to test the accuracy of atmosphere models. In
contrast, cool helium-rich stars with spectral features (i.e., DQpec,
DZ and those with CIA features) provide a real challenge to atmosphere
models and an opportunity to test our understanding of the chemistry
and physics of warm dense helium.

In this series of papers, we present and apply our new generation of
atmosphere models for cool white dwarf stars. In the first paper of
the series, we focus on improving our modeling of cool DZ
stars. Note that obtaining better fits of these objects is far more than a mere
aesthetic whim.  Indeed, because they show spectral lines, cool DZ stars
represent a unique opportunity to probe the physics and chemistry of
cool helium-rich atmospheres. In a way, they allow us to test the
models used for DC stars. Once we will have proven that our new models
are able to reproduce the rich and complex spectra of cool DZ stars,
we will be confident that the constitutive physics is accurate and
that the models can reliably be used to measure the atmospheric
parameters of all DC stars in general.

This paper describes our new model atmosphere code that includes all nonideal
effects relevant for the modelling of the atmospheres of cool DZ and DC stars.
This updated atmosphere code is based on the one described
in \cite{dufour2007spectral}. Building on other published works, as well as on
our own new calculations, we have considerably improved the
constitutive physics in our code. Section \ref{sec:opacities}
describes the additions made to correctly calculate radiative
opacities and, in Section \ref{sec:eos}, we discuss the improvements
related to the equation of state and the chemical equilibrium.  Among
the new physics added to the chemical equilibrium calculations, we
used ab initio techniques to implement a state-of-the-art description
of the chemical equilibrium of heavy elements (C, Ca, Fe, Mg and Na)
in the dense atmosphere of cool DZ stars. These calculations are
detailed at length in Section \ref{sec:ionization_main}.  In Section
\ref{sec:applications}, we present two applications that show how the
improvements included in our models translate in terms of
spectroscopic fits.  Finally, in Section \ref{sec:conclusion}, we
summarize our results and outline the upcoming papers of this series.

\section{Radiative opacities}
\label{sec:opacities}

In this Section, we describe the additions brought to the code of
\cite{dufour2007spectral} regarding the calculation of radiative
opacities. This includes improved line profiles, high-density CIA
distortion and continuum opacities corrected for collective
interactions.

\subsection{Line profiles}
\label{sec:line_profiles}
In the atmosphere of cool DZ stars, the wings of heavy element absorption lines are
severely broadened by interactions with neutral
helium. Hence, Lorentzian profiles poorly reproduce observed spectral features.
It is thus an absolute necessity to implement the unified line shape
theory described in \cite{allard1999effect} to treat such line
profiles. We implemented this formalism for the strongest
transitions found in cool DZ white dwarfs (see Table
\ref{tab:lines_allard}). In particular, the line profiles described in
\cite{allard2014caii}, \cite{allard2014nai}, \cite{allard2016mgii}, \cite{allard2016mgi} and
Allard et al. (in prep.) are used to compute the wings and a conventional
Lorentzian profile is assumed for the core of spectral lines, where the
density is low enough for this approximation to hold. To connect the two profiles,
we use a hyperbolic tangent function, which allows a smooth transition.
It should also be noted that our \ion{Ca}{1} 4226\,{\rm \AA} profile is still
preliminary, as we do not yet have access to the high-quality ab initio potentials
required for the computation of this particular line profile. To make up for this lack,
we computed our own ab initio potentials through
open-shell configuration-interaction singles calculations with the ROCIS
module of the ORCA quantum chemistry package\footnote{\url{https://orcaforum.cec.mpg.de}}
\citep{neese2012orca}.

For transitions not listed in Table \ref{tab:lines_allard}, our code assumes 
a simple Lorentzian function or quasistatic van der Waals broadening (Koester priv. comm.;
\citealt{walkup1984collisional}). Note that the exact treatment of
these secondary transitions has a limited impact on our atmospheric
determinations.

\begin{deluxetable}{cc}
  \tablecaption{Metal line profiles computed using the unified line shape theory 
    described in \cite{allard1999effect}. \label{tab:lines_allard}}
  \tablehead{\colhead{Lines} & \colhead{Source}}
  \startdata
  \ion{Ca}{1} 4226\,\AA & Allard, priv. comm. \\
  \ion{Ca}{2} H \& K &\cite{allard2014caii} \\
  \ion{Mg}{1} 2852\,\AA & Allard et al. (in prep.) \\
  \ion{Mg}{2} 2795/2802\,\AA & \cite{allard2016mgii} \\
  Mgb triplet & \cite{allard2016mgi} \\
  \ion{Na}{1} D doublet & \cite{allard2014nai} 
  \enddata
\end{deluxetable}

\begin{figure*}
  \includegraphics[width=\linewidth]{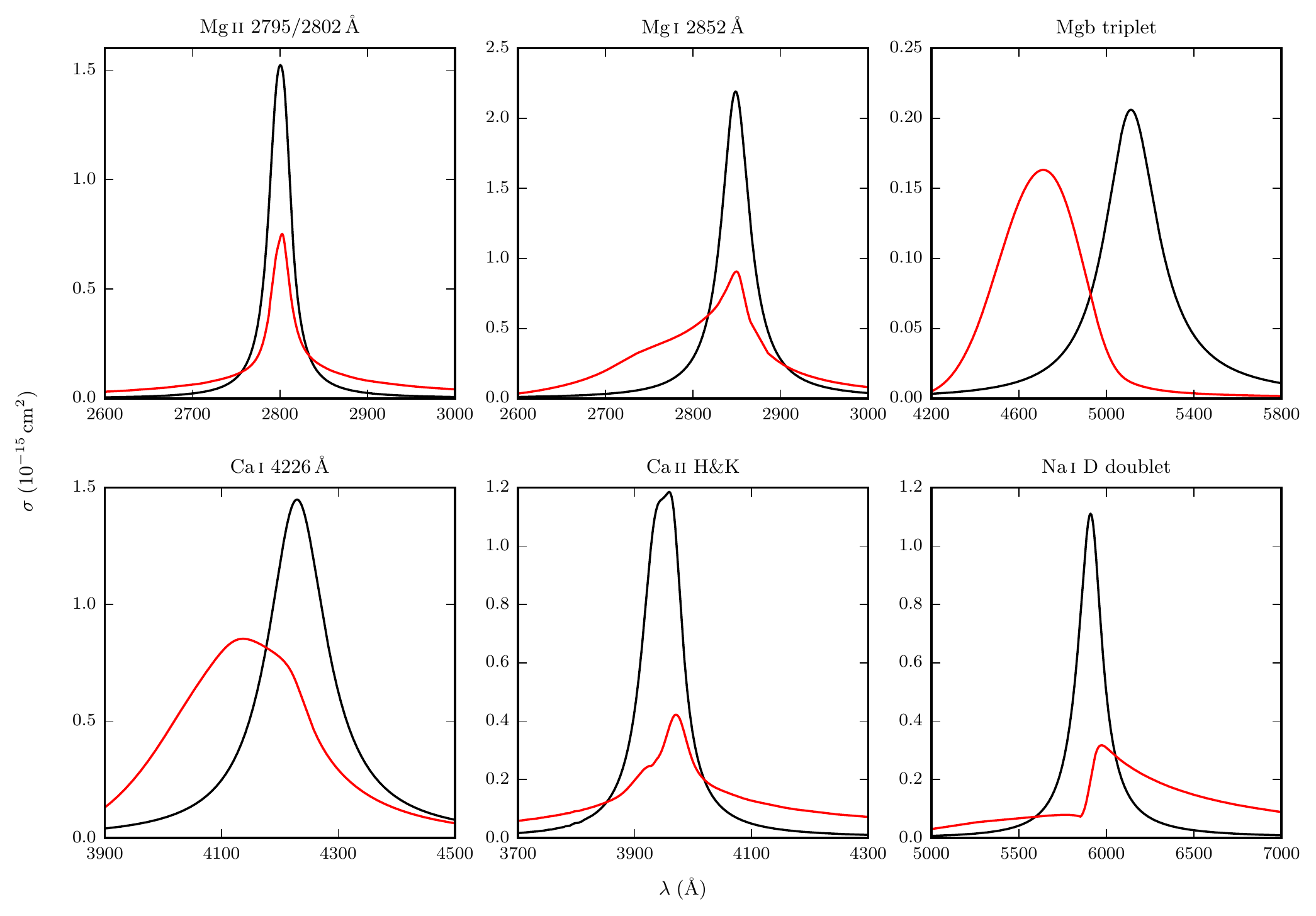}
  \caption{Absorption cross section of metal spectral lines. The black lines correspond to
    the Lorenztian profiles and the red ones are the profiles obtained with the unified line 
    shape theory of \cite{allard1999effect}. These line profiles were computed assuming 
    $T=6000\,{\rm K}$ and $n_{\rm He} = 10^{22}\,{\rm cm}^{-3}$. Note that the improved line
    profile for \ion{Ca}{1} 4226{\,\AA} relies on approximate potentials (see text).}
  \label{fig:line_prof}
\end{figure*}

We show in Figure \ref{fig:line_prof} a comparison of line profiles
calculated using the theory of \cite{allard1999effect} to those found
assuming a Lorentzian profile, for temperature and density conditions
representative of the photosphere of cool DZ stars. Clearly, under
such conditions, the Lorentzian function fails to provide a
satisfactory description of the line profiles. It underestimates the
strong broadening observed in the more accurate line profiles and does
not take into account the distortion and shift observed for many
transitions.

\subsection{Collision-induced absorption}
The calculation of the H$_2$-He CIA includes the high-density distortion effects described in 
\cite{blouin2017cia}. This pressure distortion effect alters the infrared energy distribution 
of cool DZ stars with hydrogen in their atmosphere and a photospheric density greater than 
$\approx 0.1\,{\rm g\,cm}^{-3}$ ($n_{\rm He} = 1.5 \times 10^{22}\,{\rm cm}^{-3}$). 
Moreover, we have also included the He-He-He CIA using the 
analytical fits given in \cite{kowalski2014infrared}.

\subsection{Rayleigh scattering}
In a dense helium medium, collective interactions between atoms lead to a reduction of the 
Rayleigh scattering cross section \citep{iglesias2002density}. For the wavelength domain 
relevant for white dwarf modeling (i.e., in the low-frequency limit), the reduced cross 
section can be expressed as \citep{rohrmann2018rayleigh,kowalski2006phd}
\begin{equation}
  \sigma_{\rm Rayleigh} (\omega) = S(0) \sigma_{{\rm Rayleigh}}^0 (\omega),
\end{equation}
where $\sigma_{{\rm Rayleigh}}^0 (\omega)$ is the ideal gas result \citep[e.g.,][]{dalgarno1962} 
and $S(0)$ is the structure factor of the fluid at a wavenumber $k=0$. Therefore, to take into 
account the reduction of the Rayleigh scattering, we simply need to know $S(0)$, which is a 
function of the temperature and the density of the helium fluid. To compute $S(0)$, we use the 
analytical fit to the Monte Carlo results of \cite{rohrmann2018rayleigh}.

\subsection{He$^-$ free-free absorption}
\cite{iglesias2002density} also showed that the free-free absorption cross section of the 
negative helium ion is reduced in a dense helium medium. Given that it is the dominant source 
of opacity in DZ stars, it is important to take this reduction into account. The corrected 
cross section for He$^-$ free-free absorption is given by \citep{iglesias2002density}
\begin{equation}
  \sigma_{\rm ff} (\omega) = \delta_{\rm ff} (\omega) \sigma_{{\rm ff}}^0 (\omega),
\end{equation}
where $\sigma_{{\rm ff}}^0(\omega)$ is the ideal gas result \citep[e.g.,][]{john1994absorption}. 
$\delta_{\rm ff} (\omega)$ can be computed as \citep{iglesias2002density}
\begin{equation}
  \delta_{\rm ff} (\omega) = \frac{\int_0^{\infty} I(k) {\rm d} k}{\int_0^{\infty} I_0(k) {\rm d} k},
  \label{eq:dff}
\end{equation}
where
\begin{equation}
  I(k) = I_0 (k) \frac{S(k)}{|\epsilon(\omega)|^2}
  \label{eq:Ik}
\end{equation}
and
\begin{equation}
  \begin{split}
    I_0(k) = &\frac{1}{k} \exp \left[- \frac{\hbar^2}{2m_e k_B T} 
      \left( \frac{k}{2} - \frac{m_e \omega}{\hbar k} \right)^2 \right] \\
    &\times \left| \frac{k^2 \mathcal{F} \left[\phi_{\rm e-He}(r)\right]}{4 \pi e^2} \right|^2.
  \end{split}
\end{equation}
In the last expressions, $\epsilon(\omega)$ is the dielectric function,
$m_e$ and $e$ are the electron
mass and charge, $k_B$ is the Boltzmann constant, $\hbar$ is the
reduced Planck constant and $\mathcal{F} \left[ \phi_{\rm e-He}(r)
  \right]$ is the Fourier transform of the electron-helium potential,
for which we use the simple form given by Equations 3.5 and 3.6 of
\cite{iglesias2002density}. From these equations, it follows that
two external inputs are needed to compute $\delta_{\rm ff} (\omega)$:
(1) the structure factor
$S(k)$, and (2) the index of refraction of helium
$n(\omega)=\sqrt{\epsilon(\omega)}$. The details regarding the
calculation of the structure factor are given below, while our
evaluation of the index of refraction is described in Section
\ref{sec:refraction}.

\begin{figure}
  \includegraphics[width=\columnwidth]{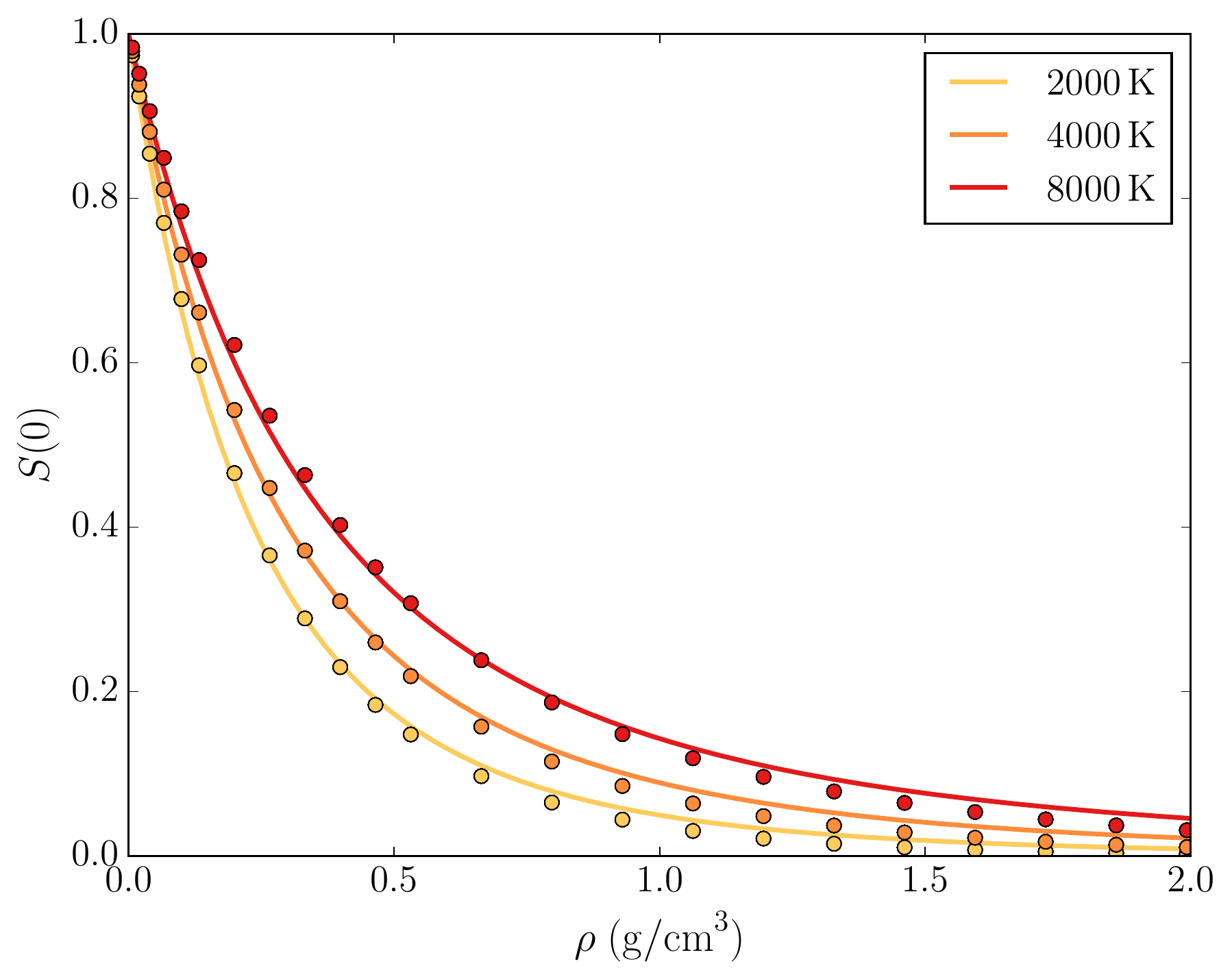}
  \caption{Structure factor at $k=0$ as a function of density and for different temperatures. 
    The solid lines show the analytical fits obtained by \cite{rohrmann2018rayleigh} from Monte 
    Carlo calculations and the circles show the results we found by solving the OZ equation.}
  \label{fig:S0_comp}
\end{figure}

To compute $S(k)$, we rely on the classical fluid theory and the
Ornstein-Zernike (OZ) equation. To solve the OZ equation, we use the
Percus-Yevick closure relation \citep{percus1958analysis}, since it is
well-suited for fluids dominated by short-range interactions (i.e.,
non-coulombic interactions; \citealt{hansen2006theory}). The
calculations are performed using a modified version of
pyOZ\footnote{\url{http://pyoz.vrbka.net}}.  Figure \ref{fig:S0_comp}
compares our $S(0)$ values to the $S(0)$ analytical fit given in
\cite{rohrmann2018rayleigh}. The agreement between both datasets is
satisfactory under $\rho = 1\,{\rm g\,cm}^{-3}$ ($n_{\rm He} = 1.5
\times 10^{23}\,{\rm cm}^{-3}$), but worsens at higher
densities. This disagreement reflects the limitations of the
Percus-Yevick closure relation at high densities, in a regime where
the Monte Carlo calculations of \cite{rohrmann2018rayleigh} are more
appropriate. Nevertheless, this small discrepancy is of limited
importance in the context of the modeling of cool DZ stars, since the
photospheric density of our models never exceeds $\approx 1\,{\rm
  g\,cm}^{-3}$.

\subsection{Index of refraction}
\label{sec:refraction}
The index of refraction, which is needed to compute the correction to the He$^-$ free-free 
cross section (Equations \ref{eq:dff} and \ref{eq:Ik}), is obtained from the Lorentz-Lorenz 
equation,
\begin{equation}
  \frac{n^2-1}{n^2+2} = A_R \left( \frac{n_{\rm He} a_0 ^3}{N_A} \right) +
  B_R \left( \frac{n_{\rm He} a_0 ^3}{N_A} \right)^2 + \mathcal{O}(n_{\rm He}^3),
  \label{eq:nref}
\end{equation}
where $A_R$ and $B_R$ are the first and the second refractivity virial 
coefficients, $n_{\rm He}$ is the helium number density, $a_o$ is the Bohr radius and $N_A$ 
is the Avogadro constant. $A_R$ is proportional to the atomic polarizability $\alpha(\omega)$
and is given by
\begin{equation}
  A_R(\omega) = \frac{4 \pi N_A \alpha(\omega)}{3}.
  \label{eq:AR}
\end{equation}
To compute $A_R$, we use the helium polarizability values reported in \cite{masili2003static}. 
For the second refractivity virial coefficient, we rely on the classical statistical mechanics 
expression \citep[e.g.,][]{fernandez1999ab}
\begin{equation}
  B_R(\omega,T) = \frac{8 N_A^2 \pi^2}{3} \int_0^{\infty} \Delta \alpha_{\rm ave} (\omega,r) 
  \exp \left[ -\frac{\phi(r)}{k_B T} \right] r^2 dr,
  \label{eq:BR}
\end{equation}
where $\Delta \alpha_{\rm ave} (\omega,r)$ is the interaction-induced isotropic polarizability 
and $\phi(r)$ is the helium-helium interatomic potential. To compute 
$\Delta \alpha_{\rm ave} (\omega,r)$, we turn to the expansion
\begin{equation}
  \Delta \alpha_{\rm ave} (\omega,r) = \Delta \alpha_{\rm ave} (0,r) + 
  \omega^2 \Delta S(-4,r) + \mathcal{O}(\omega^4),
\end{equation}
where $\Delta \alpha_{\rm ave} (0,r)$ is given in
\cite{hattig1999effect} and \cite{maroulis2000computational}, and the
Cauchy moment $\Delta S(-4,r)$ is given in
\cite{hattig1999effect}. Finally, for the interaction potential
$\phi(r)$ in Equation \ref{eq:BR}, we use the effective pair potential
of \cite{ross1986helium}, which is calibrated to fit experimental data
for high-density helium.

\begin{figure}
  \includegraphics[width=\columnwidth]{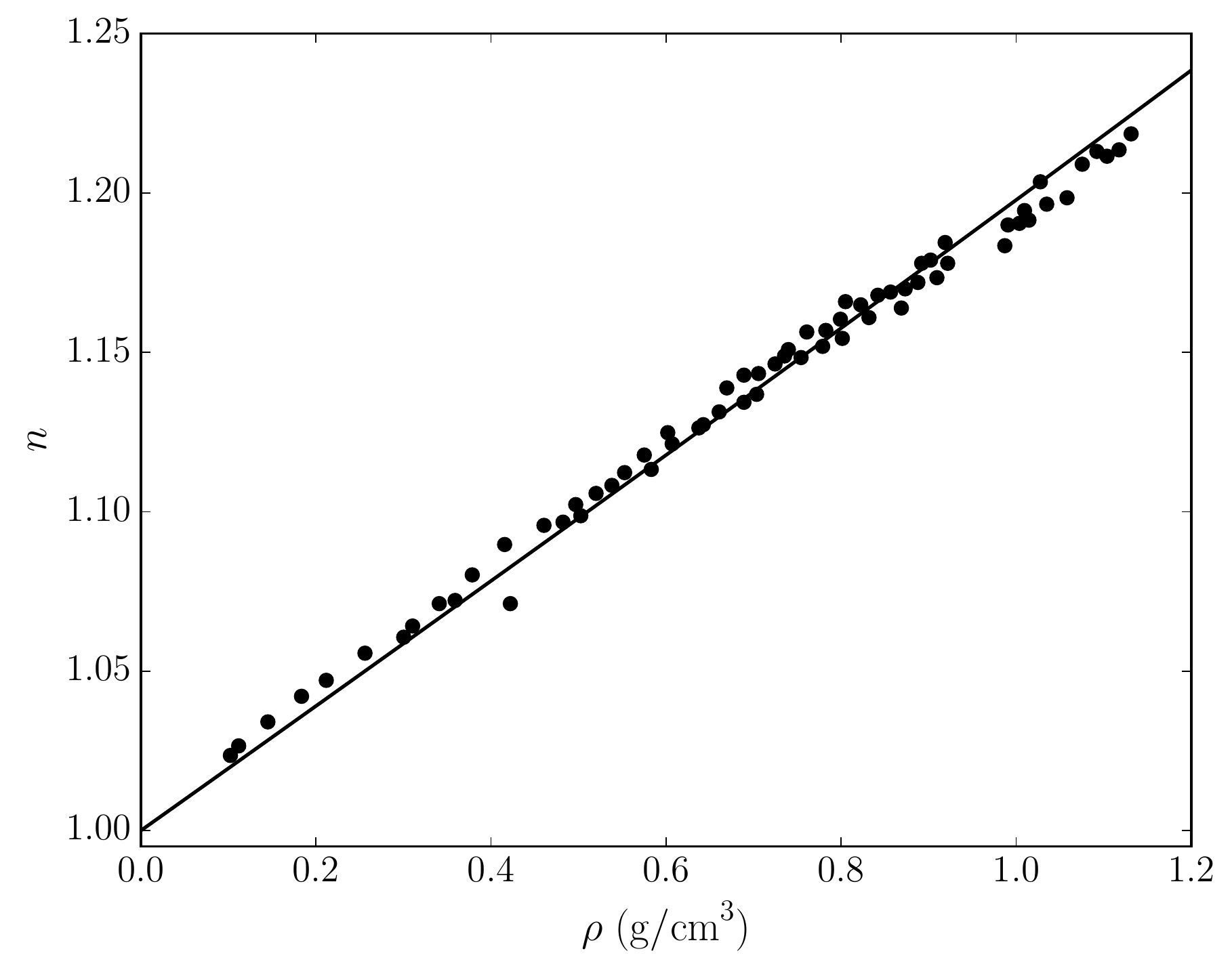}
  \caption{Index of refraction of helium as a function of density. The
    line corresponds to the results of our calculations with Equations
    \ref{eq:nref}, \ref{eq:AR} and \ref{eq:BR}, and the circles are
    the laboratory measurements extracted from
    \cite{dewaele2003measurement}.  For both datasets, $T=300\,{\rm
      K}$ and $\lambda = 6328\,$\AA.}
  \label{fig:refrac_comp}
\end{figure}

To validate our analytical model of the index of refraction, we
compared its predicted values with the high-pressure experimental
measurements of \cite{dewaele2003measurement}. This comparison
is shown in Figure \ref{fig:refrac_comp} and reveals no
significant deviation between our values and the laboratory measurements.
Additionally, we checked that our index of refraction values are 
virtually identical to those obtained by \cite{rohrmann2018rayleigh}.

\section{Equation of state and chemical equilibrium}
\label{sec:eos}
In this Section, we describe how the equation of state and the chemical equilibrium 
calculations were modified to take high-density nonideal effects into account.

\subsection{Equation of state}
The total number density and the internal energy density in each
atmospheric layer are computed using the ab initio equations of state
for hydrogen and helium published by \cite{becker2014ab}.  As in
\cite{blouin2017cia}, we resort to the additive volume rule for mixed
H/He compositions.  The mass density $\rho(P,T)$ and the internal
energy density $u(P,T)$ are given by
\begin{eqnarray}
  \frac{1}{\rho_{\mathrm{mix}}(P,T)} &=& \frac{X}{\rho_{\mathrm{H}}(P,T)} + 
  \frac{Y}{\rho_{\mathrm{He}}(P,T)}, \label{eq:rhomix} \\
  u_{\mathrm{mix}}(P,T) &=& X u_{\mathrm{H}}(P,T) + 
  Y u_{\mathrm{He}}(P,T), \label{eq:umix} 
\end{eqnarray}
where $X$ and $Y$ are the mass fractions of hydrogen and helium respectively.

\begin{figure}
  \includegraphics[width=\columnwidth]{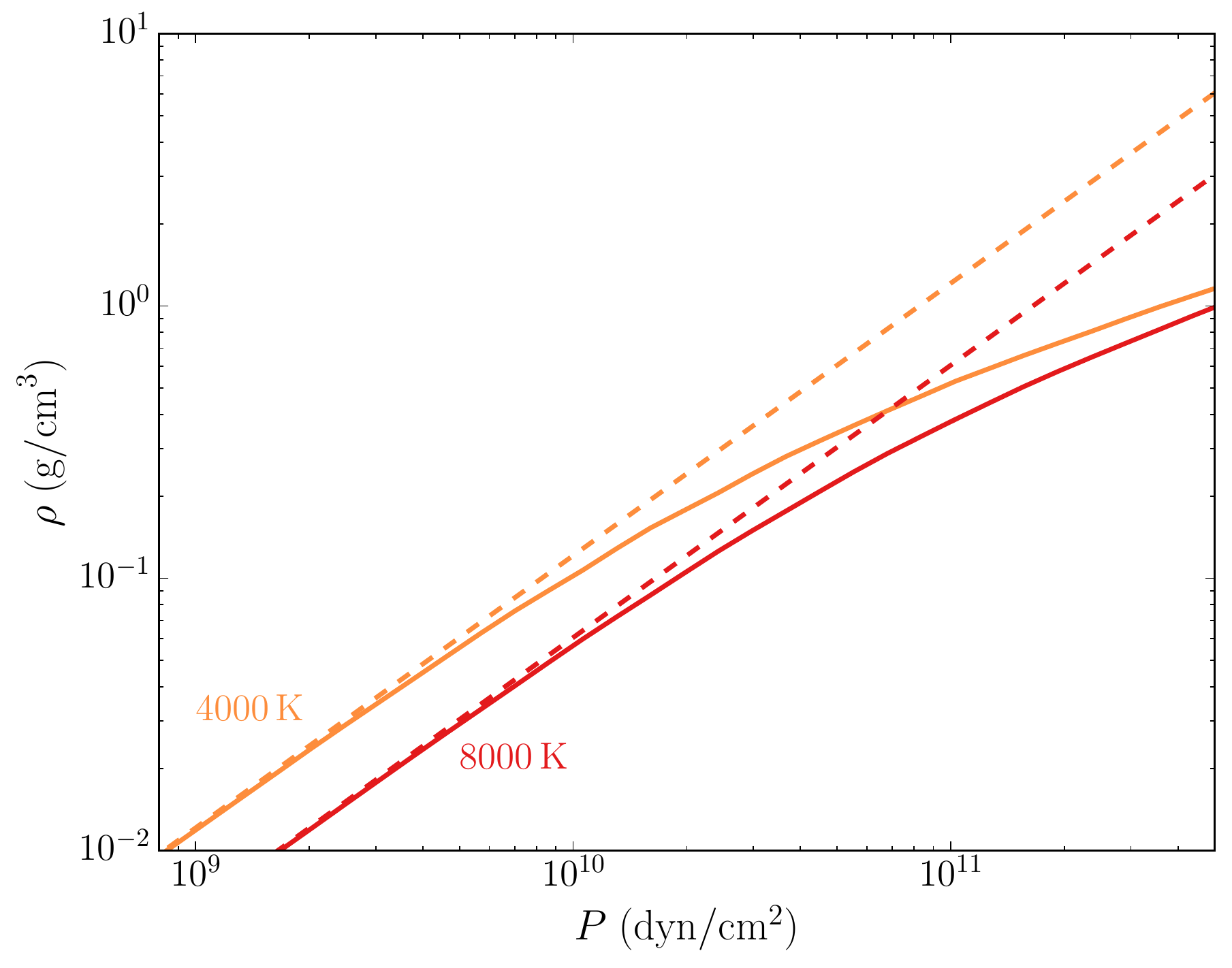}
  \caption{Density of a helium medium as a function of pressure and temperature. The solid 
    lines show the results found when using the equation of state of \cite{becker2014ab} and the 
    dashed lines correspond to the case where the ideal gas law is assumed.}
  \label{fig:nnidnid_he}
\end{figure}

For the densest cool DZ stars, the pressure at the photosphere exceeds
$10^{11}\,{\rm dyn}\,{\rm cm}^{-2}$. Under such conditions, using the
ideal gas law can lead to an important overestimation of the density.
In fact, as shown in Figure \ref{fig:nnidnid_he}, the ideal-gas
density can be up to a factor 5 greater than the value found when
using the equation of state of \cite{becker2014ab}. Such a difference
can have a significant effect on the computed atmosphere structure and
the synthetic spectrum, since most nonideal effects included in the
code (e.g., detailed line profiles, distorted CIA profiles,
high-density continuum opacities, nonideal chemical equilibrium) are
parametrized as functions of the density. For instance, using the
ideal gas law would lead to an overestimation of the broadening of
spectral lines due to an overestimation of the density of perturbing
helium atoms.

\subsection{Chemical equilibrium}
To compute the ionization equilibrium of helium, we rely on the
chemical model proposed by \cite{kowalski2007equation}. Since it does
not rely on any free parameter, this ionization equilibrium model is a
major improvement over the occupation probability formalism
\citep{hummer1988equation,mihalas1988equation2} used in most white
dwarf atmosphere codes.  Compared to models where the ideal Saha
equation is assumed, DZ models that include the helium ionization
equilibrium of \cite{kowalski2007equation} reach slightly lower
densities in their deepest layers. This is the result of pressure
ionization, which increases the electronic density and, in turn, the
opacity.  However, this effect is not as important as in metal-free
atmospheres since heavy elements provide the majority of free
electrons and therefore govern the atmosphere structure.

We have also included a detailed description of the ionization
equilibrium of heavy elements, which is the subject of Section
\ref{sec:ionization_main}.

\section{Ionization equilibrium of heavy elements}
\label{sec:ionization_main}
Properly characterizing the ionization equilibrium of heavy elements in
the atmosphere of cool DZ stars is important from several
perspectives. First, accurate ionization ratios are necessary to
obtain the right spectral line depths. For instance, in the case of a
star that shows both \ion{Ca}{2} H \& K and \ion{Ca}{1} 4226{\,\AA} in
its spectrum, obtaining the right \ion{Ca}{2}/\ion{Ca}{1} ratio is a
prerequisite for reproducing simultaneously all spectral lines.
Moreover, in cool DZ stars, heavy elements provide most of the
electrons. Therefore, a change in the ionization equilibrium of these
trace species can influence other opacity sources (most importantly
He$^-$ free-free), and hence the whole structure of the atmosphere.

Unlike the rest of the nonideal effects added to our atmosphere code,
the equilibrium of heavy elements in the dense atmosphere of cool DZ
stars has not yet been explored by other investigators using
state-of-the-art methods. Therefore, we had to perform our own
calculations before implementing this improved constitutive physics in
our code.  In this Section, we first give some theoretical background
and describe our strategy to compute the ionization equilibrium
(Section \ref{sec:theory}). Then, results from our ab initio
calculations are presented in Section \ref{sec:results} and applied to
white dwarf atmospheres in Section \ref{sec:models}.

\subsection{Theoretical framework}
\label{sec:theory}

\subsubsection{The chemical picture} 
\label{sec:chemical_picture}

To tackle the problem of the ionization equilibrium of heavy elements
in the dense atmosphere of cool white dwarfs, we rely on the chemical
picture. In this approach, atoms, ions and electrons are considered as
the basic particles and their interactions are modeled through
interaction potentials.  This is not as exact as the physical picture,
where nuclei and electrons are the basic particles.  However, using
the chemical picture has several advantages. Since this approach is
semi-analytical, the results derived from it are more easily
applicable in stellar atmosphere codes (especially regarding opacity
calculations, where thousands of bound states must be taken into
account to include the multitude of observed spectral lines). Moreover,
it is easier to identify the contribution of every physical effect and
thus gain a better physical insight of the problem at hand
\citep{winisdoerffer2005free}.

In the chemical picture, the ionization equilibrium problem
is reduced to the minimization of
the Helmholtz free energy $F(\lbrace N_i \rbrace, V, T)$ associated
with a mixture made of species $\lbrace N_i \rbrace$ in a volume $V$
maintained at temperature $T$ \citep[see for
  instance,][]{fontaine1977equations,
  magni1979thermodynamic,hummer1988equation,saumon1992fluid}. The
total Helmholtz free energy of a mixture of atoms, ions and electrons
can be expressed as the sum of the ideal free energy of the electron gas
$F^{\mathrm{id}}_{\mathrm{e}}$, the ideal free energy of every ion
from every species $F^{\mathrm{id}}_{\mathrm{j,k}}$, the contribution
from the internal structure of bound species
$F^{\mathrm{int}}_{\mathrm{j,k}}$ and the nonideal contribution related to the
interaction between species $F^{\mathrm{nid}}$,
\begin{equation}
  F = F^{\mathrm{id}}_{\mathrm{e}} + \sum_j \sum_k F^{\mathrm{id}}_{\mathrm{j,k}} + 
  \sum_j \sum_k F^{\mathrm{int}}_{\mathrm{j,k}} + F^{\mathrm{nid}},
  \label{eq:free_energy}
\end{equation}
where $k$ is an ionization state and $j$ an atomic species.

Since $F$ must be minimized, $dF=0$ and the ionization equilibrium of species $J$ between 
ionization states $K$ and $K+1$ imposes
\begin{equation}
  \begin{split}
    0 &= \left. \left( \frac{\partial F}{\partial N_e} \right|_{N_{j,k},V,T} \right) dN_e \\
    &+ \left. \left(\frac{\partial F}{\partial N_K} \right|_{N_e,N_{j,k\neq K},V,T} \right) dN_K \\
    &+ \left. \left( \frac{\partial F}{\partial N_{K+1}} \right|_{N_e,N_{j,k\neq K+1},V,T} \right) 
    dN_{K+1},
  \end{split}
\end{equation}
which, by definition of the chemical potential, is equivalent to the condition
\begin{equation}
  \mu_{J,K} = \mu_{J,K+1} + \mu_e.
  \label{eq:saha_general}
\end{equation}

Neglecting the interaction term $F^{\mathrm{nid}}$ in Equation \ref{eq:free_energy} and taking 
$F^{\mathrm{id}}_{\mathrm{e}}$ and $F^{\mathrm{id}}_{\mathrm{j,k}}$ to be the free energy of an 
ideal non-relativistic non-degenerate gas \citep{landau1980course}, Equation \ref{eq:saha_general} 
leads to the well-known Saha equation,
\begin{equation}
  \frac{n_{K+1} n_e}{n_{K}} = \frac{2 Q_{K+1}}{Q_K} \left( \frac{2 \pi m_e k_B T}{h^2} \right)^{3/2} 
  e^{-I/k_B T},
  \label{eq:saha_ideal}
\end{equation}
where $h$ is the Planck constant, $n_i$ are number densities, $Q_i$ are partition functions 
and $I$ is the ionization potential.

Now, if we keep the nonideal terms in the free energy equation, we find a result of the form of 
Equation \ref{eq:saha_ideal}, but with an \textit{effective} ionization potential $I+\Delta I$ 
\citep{kowalski2007equation,zaghloul2009thermodynamic},
\begin{equation}
  \frac{n_{K+1} n_e}{n_{K}} = \frac{2 Q_{K+1}}{Q_K} \left( \frac{2 \pi m_e k_B T}{h^2} \right)^{3/2} 
  e^{-(I+\Delta I)/k_B T},
  \label{eq:saha_nonideal}
\end{equation}
where
\begin{equation}
  \Delta I = \mu_e^{\mathrm{nid}} + \mu_{K+1}^{\mathrm{nid}} - \mu_{K}^{\mathrm{nid}}.
  \label{eq:deltaI}
\end{equation}
Therefore, to compute the nonideal ionization equilibrium of heavy elements in dense helium-rich 
fluids, all that is needed is to compute the appropriate $\Delta I$ given by the above equation.

In Equation \ref{eq:deltaI}, it is the difference in free energy of
many-body systems in thermodynamic equilibrium with different
ionization states that is computed. This yields an effective
ionization potential applicable to \textit{thermodynamic} ionization
equilibrium calculations. As emphasized by
\cite{crowley2014continuum}, this ionization potential is not directly
applicable to non-equilibrium processes (e.g., photoionization). These
are fast (adiabatic) processes that occur before the surrounding
plasma has any time to respond.

\subsubsection{General strategy}
\label{sec:strategy}
To compute $\Delta I$, we have to evaluate the nonideal chemical
potential of every species involved in the ionization process. The
electronic term $\mu_e^{\mathrm{nid}}$ is already available in the
literature. \cite{kowalski2007equation} performed density functional
theory (DFT) calculations to evaluate the excess energy of an electron
embedded in a dense helium medium and found values that are in good
agreement with existing laboratory measurements
\citep{broomall1976density}. These calculations, published as
polynomial expansions, were performed for a range of temperatures and
densities suitable for our purpose.

While $\mu_{K+1}^{\mathrm{nid}}$ and $\mu_{K}^{\mathrm{nid}}$ were
calculated by \cite{kowalski2007equation} in the case of helium
ionization, we are not aware of any study where the nonideal chemical
potentials were computed for heavy elements surrounded by dense
helium.  The central task of this Section is to compute these chemical
potentials in order to obtain $\Delta I$ by virtue of Equation
\ref{eq:deltaI}.

In the limit of strongly-coupled systems, the role of entropy can be
neglected for the calculation of thermodynamic equilibrium ionization
potential since the configuration of atoms remains the same before and
after the ionization takes place. However, plasmas encountered in
white dwarf atmospheres have a finite coupling strength. When an atom
is ionized, the medium responds and additional energy is transferred
between the atom and the surrounding particles
\citep{crowley2014continuum}. Therefore, the nonideal chemical potential
of a species in ionization state $K$ can be expressed as the sum of two
contributions,
\begin{equation}
  \mu_K^{\mathrm{nid}} = E_K^{\mathrm{exc}} + \mu_K^{\mathrm{nid,ent}},
  \label{eq:munidsplit}
\end{equation}
where $E_K^{\mathrm{exc}}$ is the excess of internal energy per particle
and $\mu_K^{\mathrm{nid,ent}}$ is the entropic contribution to the nonideal
chemical potential.
Note that this separation of $\mu_K^{\mathrm{nid}}$ into two distinct components
directly follows from the definition of the Helmholtz free energy. As $F=E+TS$
and $\mu^{\rm nid}_K=\left. (\partial F_{K}^{\rm nid} / \partial N_K) \right|_{N_{k\neq K},V,T}$, 
we can write
\begin{equation}
\mu^{\rm nid}_K = \left. \frac{\partial \left(E_{K}^{\rm nid} + TS_{K}^{\rm nid} \right)}
   {\partial N_K} \right|_{N_{k\neq K},V,T} = E_K^{\rm exc} + \mu_K^{\mathrm{nid,ent}}.
\end{equation}

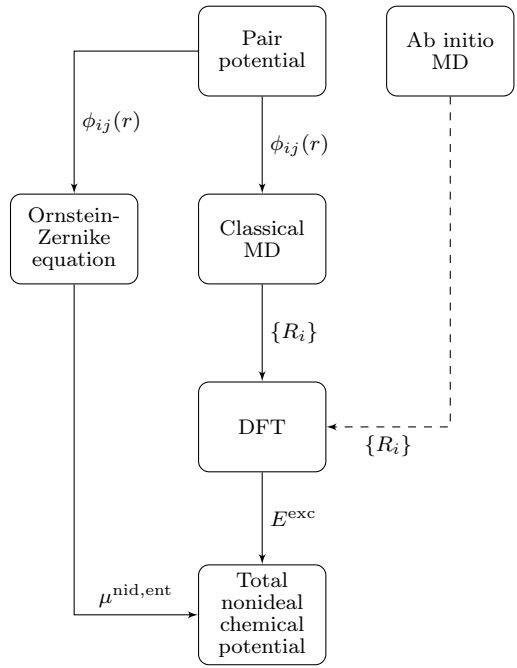
\begin{figure}
  \centering
  \tikzstyle{block1} = [rectangle, draw, fill=white, 
    text width=5em, text centered, rounded corners, minimum height=4em]
  \tikzstyle{line} = [draw, -latex']
  \begin{tikzpicture}[node distance = 2cm, auto]
    \node [block1] (start) {Pair potential};
    \node [block1, right of=start, node distance=2.5cm] (DFTMD) {Ab initio MD};
    \node [block1, below of=start, node distance=2.5cm] (MD) {Classical MD};
    \node [block1, left of=MD, node distance=2.5cm] (OZ) {Ornstein-Zernike equation};
    \node [block1, below of=MD, node distance=2.5cm] (QE) {DFT};
    \node [block1, below of=QE, node distance=2.5cm] (stop) {Total nonideal chemical potential};
    \path [line] (start) -- node {$\phi_{ij}(r)$} (MD);
    \path [line] (MD) -- node {$\lbrace R_i \rbrace$} (QE);
    \path [line] (QE) -- node {$E^{\mathrm{exc}}$}  (stop);
    \path [line] (OZ) |- node [near end] {$\mu^{\mathrm{nid,ent}}$} (stop);
    \path [line] (start) -| node [near end] {$\phi_{ij}(r)$} (OZ);
    \path [line,dashed] (DFTMD) |- node [near end] {$\lbrace R_i \rbrace$} (QE);
  \end{tikzpicture}
  \caption{Computational strategy used to retrieve the nonideal chemical potential of ionic 
    species. The dashed arrow indicates a validation step described in Section 
    \ref{sec:interaction_energy}.}
  \label{fig:strategy}
\end{figure}

Our general strategy is summarized in Figure \ref{fig:strategy}. To
compute the $\mu_K^{\mathrm{nid,ent}}$ contribution, we follow the
work of \cite{kowalski2006dissociation} and
\cite{kowalski2007equation} and use the classical fluid theory and the
OZ equation, as detailed in Section \ref{sec:excluded_volume}. To
retrieve $E_K^{\mathrm{exc}}$, we turn to DFT to compute the
excess energy of a metallic ion embedded in a dense helium
medium. This approach has the advantage of naturally taking into
account many-body interaction terms. Prior to using DFT to compute
$E_K^{\mathrm{exc}}$, we use molecular dynamics (MD) simulations
to obtain representative atomic configurations, as described in detail
in Section \ref{sec:interaction_energy}.

\subsubsection{Comparison with previous studies}

To take into account the nonideal ionization of heavy elements, white dwarf
atmosphere models
\citep{dufour2007spectral,koester2000element,wolff2002element}
typically rely on the Hummer-Mihalas occupation probability
formalism \citep{hummer1988equation,mihalas1988equation2}. In this
framework, an occupation probability $w_i$ is assigned to every
electronic level of every ion.  If the level is unperturbed, $w_i=1$;
if the level is completely destroyed by interparticle interactions,
$w_i=0$. This occupation probability appears in the Boltzmann
distribution and it multiplies every term of the partition function,
\begin{equation}
  Q_K = \sum_i w_{iK} g_{iK} \exp \left( -\frac{e_{iK}}{k_B T} \right),
  \label{eq:partHM}
\end{equation}
where the sum is over all states $i$ of species $K$, and $g$ is a statistical weight. To compute 
$w_i$ in the particular case of neutral interactions, \cite{hummer1988equation} use the second 
virial coefficient in the van der Waals equation of state to obtain
\begin{equation}
  w_i = \exp \left[ - \frac{4 \pi}{3} \sum_{i'} n_{i'} (r_i + r_{i'} )^3 \right],
  \label{eq:hm}
\end{equation}
where $n_i$ is the number density of particles in state $i$ and $r_i$
is the radius of the particles in this state. The interpretation of
Equation \ref{eq:hm} is straightforward: when a state occupies a
volume of the same order as the mean volume allowed per particle, it
is gradually destroyed. Although simple and easy to implement in
atmosphere models, we see three important drawbacks with this
approach.
\begin{enumerate}
\item This formalism is expected to break down above $\approx 0.01\,{\rm g\,cm}^{-3}$ 
  \citep{hummer1988equation}, which is insufficient for many cool DZ white dwarfs.
\item The excluded volume effect is only a caricature of the real interaction potential between 
  two neutral particles.
\item There is no theoretical prescription for the radii $r_i$. For instance, for a ground 
  state \ion{He}{1} atom, should $r$ be given by the hydrogenic approximation 
  ($r= n^2 a_0 / Z_{\mathrm{eff}}=0.39\,$\AA) or should it be given by the van der Waals 
  radius \citep[$1.40\,$\AA,][]{bondi1964van}? To address this problem, it is always possible 
  to calibrate the radii to fit the spectral lines observed in white dwarf stars. This was 
  successfully done by \cite{bergeron1991synthetic} for hydrogen, but it would be impracticable 
  for DZ stars, where many ions contribute to the total electronic density.
\end{enumerate}

Our approach aims at answering these three concerns. First, by taking
into account many-body interaction terms, it is designed to remain
physically exact up to densities of the order of $1\,{\rm g\,cm}^{-3}$.
Secondly, the interaction between species is modeled through ab
initio calculations that accurately describe the complex behavior of
electrons under these high-density conditions. Finally, since we rely
only on first-principles physics, our method does not require any free
parameter.

\subsubsection{Approximations}
\label{sec:approximations}
Before moving to the calculation of the nonideal chemical potentials and $\Delta I$ in Section 
\ref{sec:results}, we take time to justify three important approximations that we use throughout 
Section \ref{sec:ionization_main}.

\paragraph{Electrons and heavy elements as trace species}
\label{sec:trace_approx}
We are interested in helium-rich plasmas, where heavy elements and
electrons can be considered as trace species. Hence, we completely
neglect the interaction of metallic ions with other metallic ions and
with electrons. This approximation is justified by the very low
abundance of heavy elements in white dwarf atmospheres. Indeed, to our
knowledge, the most metal-rich DZ star mentioned in the literature has
an atmosphere with a number density ratio of $\log \mathrm{Ca} /
\mathrm{He} \approx -6$ \citep[Ton 345,][]{wilson2015composition}.

\begin{figure}
  \centering
  \includegraphics[width=\columnwidth]{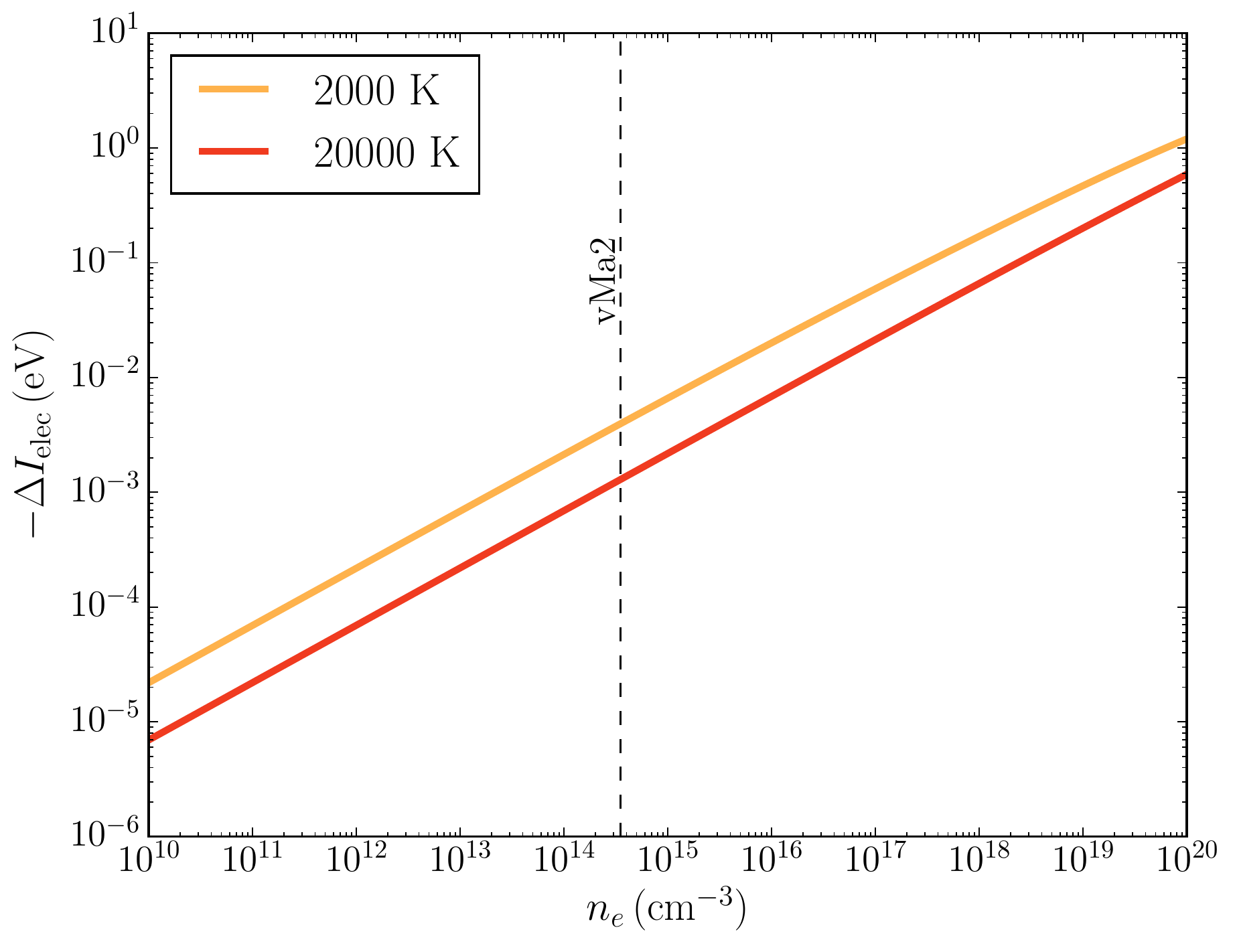}
  \caption{Contribution of the electrostatic interaction to the effective ionization potential 
    with respect to the electronic density and the temperature. The dashed line indicates the 
    electronic density at $\tau_R = 2/3$ for vMa2, a typical cool DZ star.}
  \label{fig:muelec}
\end{figure}

As a consequence of this approximation, we completely ignore the
excess energy resulting from the interaction between charged
species. Since electrostatic interactions occur at long range, this
approximation deserves some additional justifications. To show that
electrostatic interactions are negligible, we computed the
contribution of electrostatic interactions to the Helmholtz free
energy. The latter can be broken down into three components
\citep{chabrier1998equation},
\begin{equation}
  F^{\mathrm{elec}} = F^{ee} + F^{ii} + F^{ie},
\end{equation}
where $F_{ee}$ is the exchange-correlation contribution from the
electron fluid, $F^{ii}$ is the contribution from the one-component
ion plasma and $F^{ie}$ is the electron screening contribution. To
evaluate $F^{\mathrm{elec}}$, we used the equations reported in
\cite{ichimaru1987statistical} for $F^{ee}$ and those in
\cite{chabrier1998equation} for $F^{ii}$ and $F^{ie}$. If all
electrons originate from singly-ionized species, then
$F^{\mathrm{elec}}$ is a function of only the electronic density $n_e$
and the plasma temperature $T$. Figure \ref{fig:muelec} shows $\Delta
I^{\mathrm{elec}} = \left( \frac{\partial}{\partial N_e} +
\frac{\partial}{\partial N_{j,i+1}} \right) F^{\mathrm{elec}}$ for
different $n_e$ and $T$. The dashed line indicates the electronic
density at the photosphere ($\tau_R = 2/3$) of vMa2, a typical cool DZ
star. At these electronic densities and temperatures, the effect of
electrostatic interactions on $\Delta I$ is of only a few meV and is
therefore negligible compared to the total $\Delta I$ reported later
in this paper (which is of the order of a few eV). The charged
particles density is simply too low for electrostatic interactions to
have any significant effect.

\paragraph{Omission of the quantum behavior of ions}
\label{sec:quantum_corr}
We do not take into account the quantum behavior of ions and atoms. To
justify this approximation, we can compute the first quantum
correction of the Helmholtz free energy \citep{wigner1932quantum},
which can be seen as a correction for the overlapping wave functions
of nearby particles. For an $m$-component mixture, it can be expressed
as \citep{saumon1991fluid}
\begin{equation}
  F^{\mathrm{quant}} = \frac{\pi \hbar^2}{12 k T V} \sum_{a,b}^{m} \frac{N_a N_b}{\mu_{ab}} 
  \int \nabla^2 \phi_{ab} (r) g_{ab}(r) r^2 dr,
\end{equation}
where $\phi_{ab} (r)$ and $g_{ab}(r)$ are, respectively, the pair
potential and the pair distribution function between species $a$ and
$b$, and $\mu_{ab} = \frac{m_a m_b}{m_a + m_b}$ is the reduced mass of
particles $a$ and $b$. The contribution of this term to $\Delta I$ is
computed as $\Delta I^{\mathrm{quant}} = \left(
\frac{\partial}{\partial N_{j,k+1}} - \frac{\partial}{\partial
  N_{j,k}} \right) F^{\mathrm{quant}}$. Using the pair distribution
functions and the pair potentials described in Section
\ref{sec:excluded_volume}, we find that $\Delta I^{\mathrm{quant}}$
remains below $5\,\mathrm{meV}$ for all physical conditions relevant
for the modeling of the atmosphere of cool DZ stars. As this is well
below $E^{\mathrm{exc}}$ and $\mu^{\mathrm{nid,ent}}$, we can
safely ignore the quantum behavior of ions.

\paragraph{The ground-state approximation}
\label{sec:ground_state_approx}
To compute the ionization equilibrium of heavy elements, we assume that every atom is in its 
electronic ground state. This solely means that we consider all species to be in their ground 
state \textit{when computing the ionization equilibrium}. Once the ionization equilibrium is
computed, the population of every electronic state can be obtained through the Boltzmann 
distribution. How good is this approximation? For helium atoms, this approximation is excellent. 
The first excited state of \ion{He}{1} lies at 19.8\,eV, so almost all helium atoms are in their 
fundamental state for the temperature domain in which we are interested ($k_B T<1\,$eV). 

For heavy elements, this approximation could be problematic. It is
well known that excited states are typically more affected by nonideal
effects than the fundamental state
\citep[e.g.,][]{hummer1988equation}. Therefore, since the $\Delta I$
term in Equation \ref{eq:saha_nonideal} only takes into account the
destruction of the fundamental state, an error could be introduced in
the ionization equilibrium if excited states are affected in a
significantly different way \textit{and} if they account for a large
portion of the partition function $Q$.

\begin{figure}
\centering
\includegraphics[width=\columnwidth]{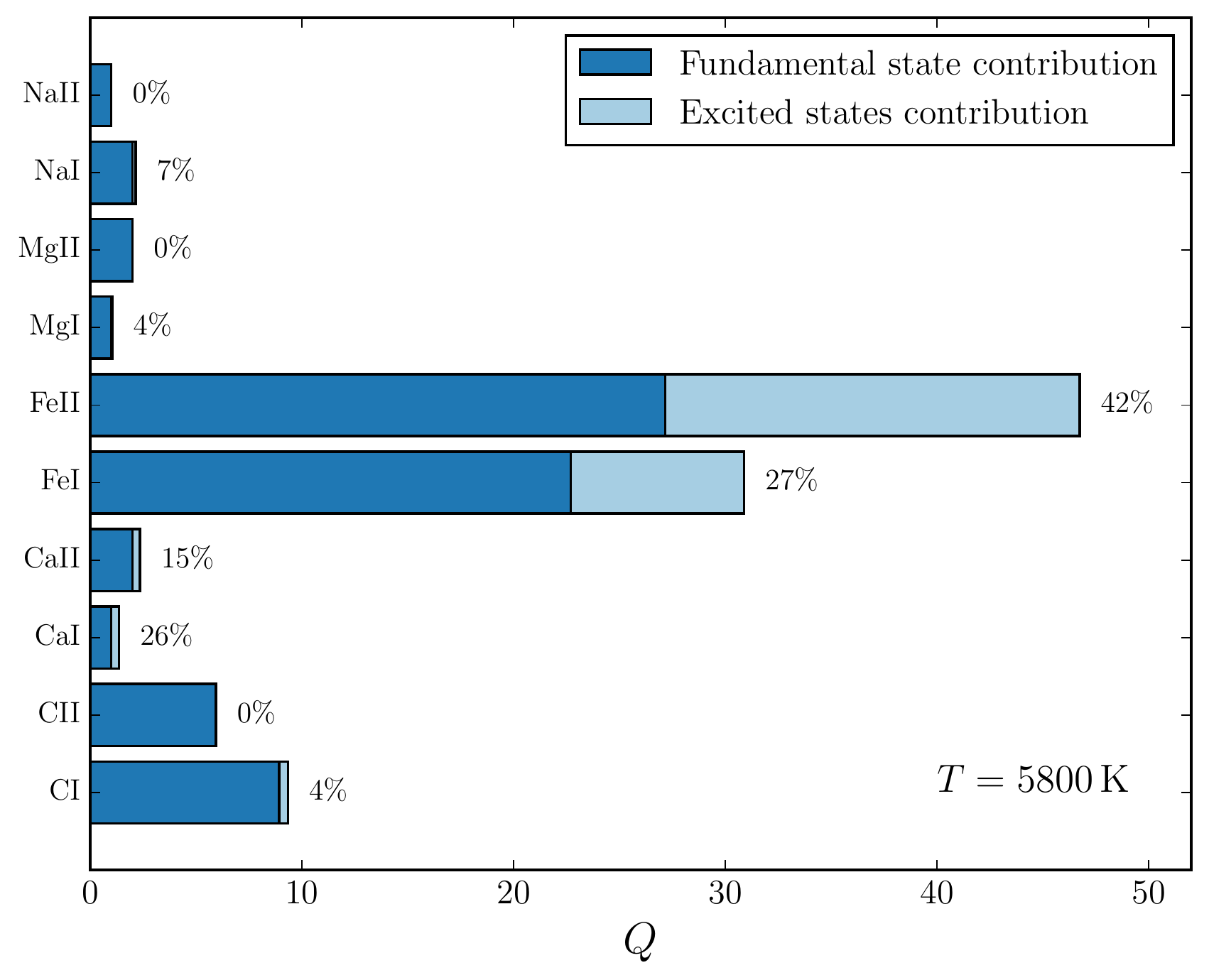}
\caption{Comparison of the contributions of the fundamental state and the excited states to the 
  partition function $Q$ at \mbox{$k_B T=0.5\,$eV} for heavy ions found in cool DZ stars. 
  The number at the end of each bar gives the fraction of $Q$ resulting from excited states. 
  This figure was made using the atomic data of the NIST Atomic Spectra Database \citep{NIST_ASD}.}
\label{fig:Qcontrib}
\end{figure}

To investigate the maximum error associated with this approximation,
we computed the fundamental state contribution to the partition
function $Q$ for C, Ca, Fe, Mg and Na.  The results are shown in
\mbox{Figure \ref{fig:Qcontrib}} for $k_B T=0.5\,$eV. The worst
possible error associated with this approximation will occur if all
excited states are destroyed while the fundamental state remains
unperturbed (see Equation \ref{eq:partHM}). This scenario is highly unlikely,
but provides an easy way of assessing the maximum error. If it is the
case, then, as shown in \mbox{Figure \ref{fig:Qcontrib}}, the maximum
error on $Q$ is $\approx 40\%$ (see \ion{Fe}{2}).  Therefore, in the
worst case, the ionization fraction will be wrong by a factor of
$\approx 2$.

This maximum error is not a cause of concern for the modeling of the
atmosphere of cool DZ stars.  First, for all other atomic species (C,
Ca, Mg and Na) $Q$ is far more dominated by the fundamental state
contribution and the maximal error associated with this approximation
is thus much smaller than the value derived for Fe. Secondly, for the
coolest DZ stars, the relative contribution of the fundamental state
to the partition function is higher than for their warmer
counterparts. Therefore, the ground-sate approximation becomes more
accurate for the stars for which the departure for the ideal chemical
equilibrium is expected to be the most important. Last but not least,
for the conditions relevant for the modeling of cool DZ stars, both
this work and the formalism of \cite{hummer1988equation} predict
deviations for the ideal gas equilibrium that are much more important
than the aforementioned factor of $\approx 2$ (see for instance Figure
\ref{fig:comp_hm}).

\subsection{Results}
\label{sec:results}
In this section, we detail the computations performed to obtain
$\Delta I$ for C, Ca, Fe, Mg and Na. In Sections
\ref{sec:excluded_volume} and \ref{sec:interaction_energy}, we
describe the computational setup and our intermediate results, and our
final results are given in Section \ref{sec:final_results}. For the
sake of clarity, we only refer to Ca in
the discussion of Sections \ref{sec:excluded_volume} and
\ref{sec:interaction_energy}, although all the reported calculations
were also performed for C, Fe, Mg and Na.

\subsubsection{Entropic contribution}
\label{sec:excluded_volume}

To compute the entropic contribution to the nonideal chemical
potential, we first use the OZ equation (and the Percus-Yevick closure
relation) to find the radial distribution function
$g_{\mathrm{He-Ca}}(r)$ describing the spatial configuration of Ca
relative to He atoms.  Then, once the radial distribution function
$g_{\mathrm{He-Ca}}(r)$ is obtained,
$\mu_{\rm Ca}^{\rm nid}$ can be obtained through
Equations 9 and 12 of \cite{kiselyov1990free}. From there, we simply substract
the excess energy of Ca (as computed in the OZ framework) to obtain
$\mu_{\rm Ca}^{\rm nid,ent}$ (Equation \ref{eq:munidsplit}).

To compute $g_{\mathrm{He-Ca}}(r)$ with the OZ equation,
the pair potentials $\phi_{\mathrm{He-He}}(r)$ and
$\phi_{\mathrm{He-Ca}}(r)$ must be specified (in accordance with the
approximation detailed in Section \ref{sec:approximations},
$\phi_{\mathrm{Ca-Ca}}(r)=0$ since the metal-metal interactions are
neglected).  For the helium-helium pair potential, we use the
effective pair potential of \cite{ross1986helium}.

As metal-helium pair potentials are not available in the literature
for every metallic ion considered in this work, we had to compute ab
initio pair potentials between helium and metallic ions. To do so, we
used the ORCA quantum chemistry
package to obtain the potential energy
$\phi_{\mathrm{Ca-He}}$ at various separations,
\begin{equation}
  \phi_{\mathrm{Ca-He}} (r) = E_{\mathrm{Ca-He}} (r) - E_{\mathrm{He}} - E_{\mathrm{Ca}},
\end{equation}
where $E_{\mathrm{Ca-He}} (r)$ is the total energy for a separation $r$ and $E_{\mathrm{He}}$ 
and $E_{\mathrm{Ca}}$ are the computed energies of isolated He and Ca atoms. We rely on the 
CCSD(T) method \citep{raghavachari1989fifth} as implemented in ORCA 
\citep{kollmar2010coupled,neese2009accurate} with the \mbox{aug-cc-pCVQZ} basis sets 
\citep{dunning1989gaussian,kendall1992electron,woon1993gaussian}. Using the counterpoise method 
\citep{boys1970calculation}, we verified that the basis set superposition error is small enough 
($<2\,$meV) to be neglected for our purpose.

\begin{figure}
  \centering
  \includegraphics[width=\columnwidth]{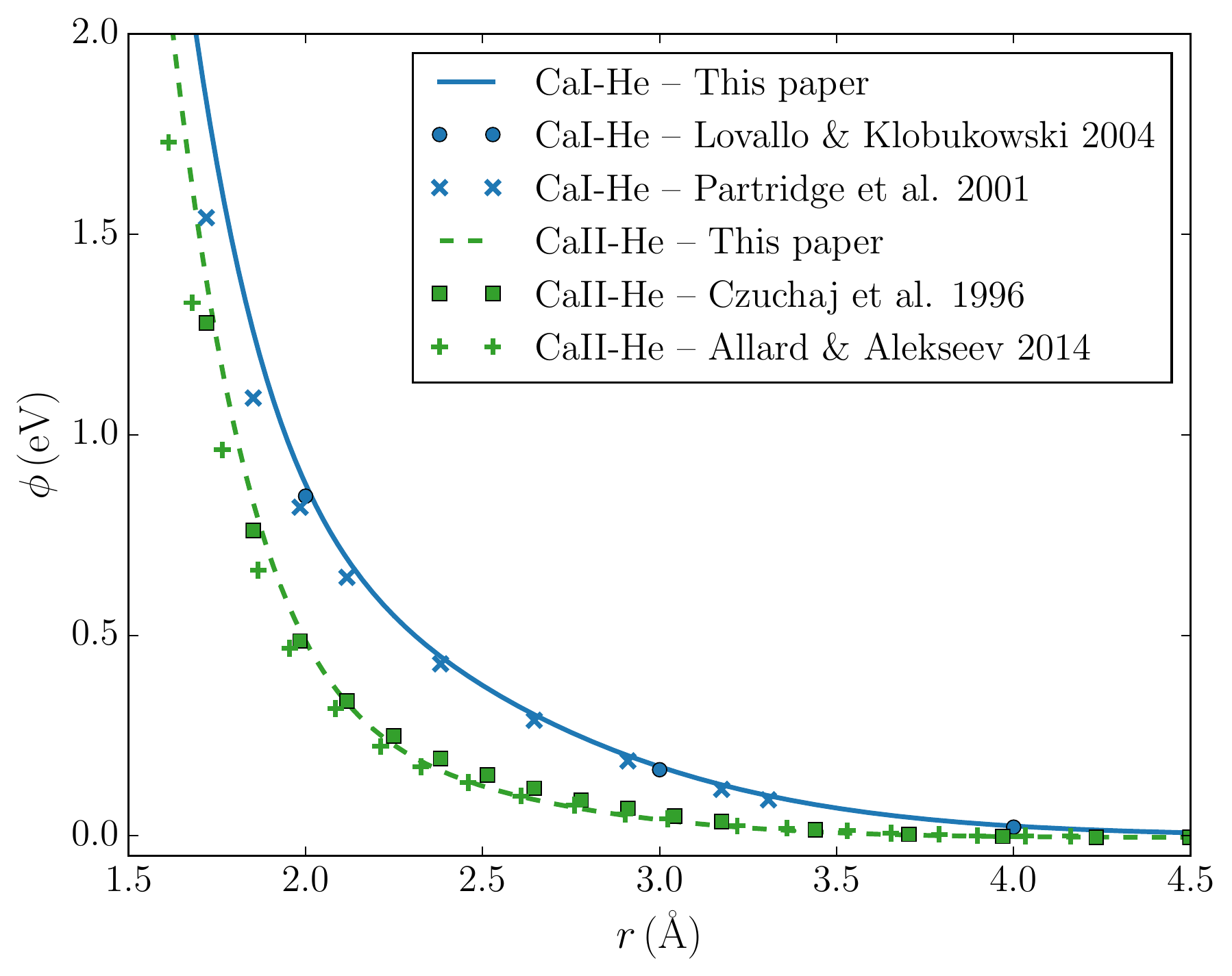}
  \caption{Comparison between the pair potentials for the \ion{Ca}{1}--\ion{He}{1} and 
    \ion{Ca}{2}--\ion{He}{1} interactions computed in this work and the values reported in 
    \cite{lovallo2004accurate}, \cite{partridge2001potential}, \cite{czuchaj1996pseudopotential} 
    and \cite{allard2014caii}}
  \label{fig:comp_cahe}
\end{figure}

In the particular case of Ca, a few interaction potentials can be
found in the literature for the \ion{Ca}{1}--\ion{He}{1}
\citep{lovallo2004accurate,partridge2001potential} and the \ion{Ca}{2}--\ion{He}{1}
interactions
\citep{allard2014caii,czuchaj1996pseudopotential}.  We used the
values reported by these authors to validate our computational
setup. This comparison, which reveals no significant differences, is
shown in Figure \ref{fig:comp_cahe}.

The main limitation of these pair potentials is that they were
obtained in the infinite-dilution limit (i.e., Ca interacts with only
one He atom). Therefore, when we use these potentials, we implicitly
assume that the total potential is pairwise additive, and an
error may be introduced if many-body terms are
important. This is the main reason why we resort to the OZ equation
only to compute the entropic contribution and not to compute
the excess energies. In fact, as described in Section
\ref{sec:interaction_energy}, we turn to DFT to compute excess
energies, which guarantees that many-body interaction terms are
properly taken into account.

\subsubsection{Excess energy contribution}
\label{sec:interaction_energy}
The excess energy of Ca embedded in a dense helium medium made of $N$ He atoms is given by
\begin{equation}
  E_{\mathrm{Ca-He}}^{\mathrm{exc}} = E_{N\mathrm{He}+\mathrm{Ca}} - E_{N\mathrm{He}} - E_{\mathrm{Ca}},
  \label{eq:ecahe_int}
\end{equation}
where $E_{N\mathrm{Ca}+\mathrm{He}}$ is the total energy of the system, $E_{N\mathrm{He}}$ is the 
energy of the $N$ He atoms and $E_{\mathrm{Ca}}$ is the computed energy of the isolated Ca atom. 
This calculation requires two steps. First, we need to find meaningful atomic configurations for
the system (i.e., configurations that are representative of the thermodynamic fluctuations 
undergone by the real system). Then, we can use these configurations to compute the excess
energy with Equation \ref{eq:ecahe_int}.

\paragraph{Molecular dynamics}
\label{sec:molecular_dynamics}
To obtain representative atomic configurations of a system consisting
of one Ca atom surrounded by $N$ He atoms at a given temperature and a
given density, we turned to classical molecular dynamics
simulations. More precisely, we used
LAMMPS\footnote{\url{http://lammps.sandia.gov}}
\citep{plimpton1995fast} and the pair potentials described in Section
\ref{sec:excluded_volume}.  The simulations were performed in a cubic
box with periodic boundary conditions. The box size and the number of
He atoms included in the simulations were chosen to attain the desired
density (additional considerations regarding finite-size effects are
discussed in the next paragraph) and the temperature was kept near the
target value using a Nos{\'e}-Hoover thermostat
\citep{nose1984unified,hoover1985canonical}. The simulations were run
for 5\,ns using 0.2\,fs time steps. At regular time intervals, the
atomic positions were saved and it is these configurations that we use
in the next Section to compute the excess energies.

\paragraph{DFT calculations}
\label{sec:DFT_calculations}

To compute the excess energy of Ca in the atomic configurations
extracted from the molecular dynamics simulations, we used the {\sc
  Quantum ESPRESSO}\footnote{\url{http://quantum-espresso.org}} DFT
package \citep{giannozzi2009quantum}, with the PBE
exchange-correlation functional \citep{perdew1996generalized} and
norm-conserving pseudopotentials. For all DFT calculations, we chose a
kinetic energy cutoff of \mbox{45\,Ry} \mbox{(612\,eV)} and a charge density cutoff of
\mbox{180\,Ry}.  We checked that this cutoff is enough to achieve a
$<0.05\,$eV convergence of the metal excess energy. To remove the
electrostatic interaction associated with periodic boundary
conditions, we used the Martyna-Tuckerman correction
\citep{martyna1999reciprocal} as implemented in {\sc Quantum
  ESPRESSO}, which allows to correct both the total energy and the SCF
potential.

Furthermore, to make sure that the finite size of the box does not
result in undesired artifacts, we performed simulations using
different numbers of helium atoms per simulation box and different box
sizes (up to $N=160$ helium atoms and up to $a=30\,$a.u.). We found
that using at least $N=50$ helium atoms and a simulation box of at
least $a=15$\,a.u. (7.94{\,\AA})
allows a $<0.1\,$eV convergence of the excess
energy compared to results obtained at the same density with higher
$N$ and $a$ values.  This indicates that finite-size artifacts are negligible
when these two conditions are met.  Hence, all DFT calculations
reported in this work were performed with $a\geq 15$\,a.u. and $N\geq
50$.

\begin{figure}
  \centering
  \includegraphics[width=\columnwidth]{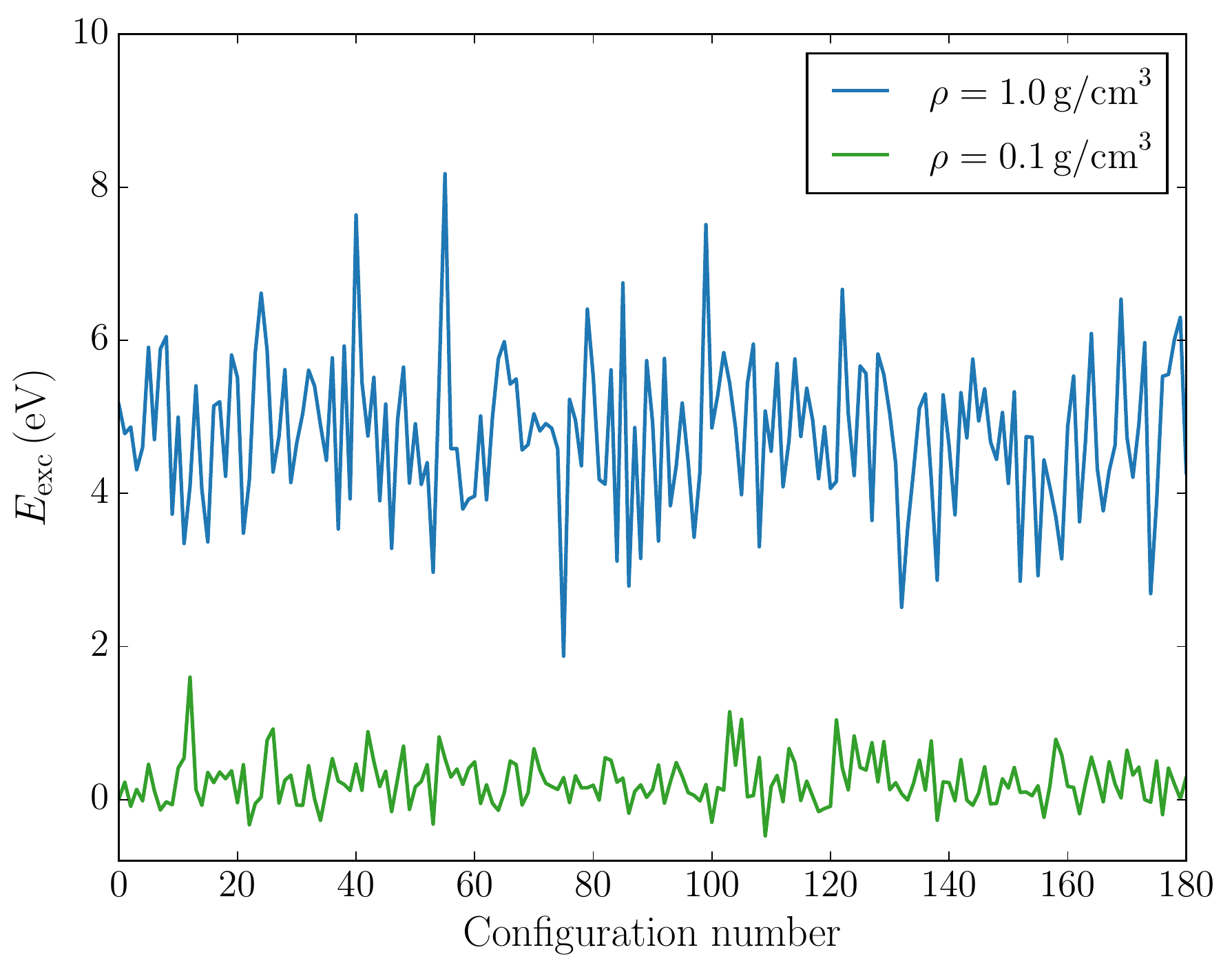}
  \caption{Excess energy of Ca at $T=4000\,$K for 
    configurations taken at $25\,$ps intervals from MD trajectories, for
    different helium densities.}
  \label{fig:eexc_vs_nsnap}
\end{figure}

\begin{figure}
  \centering
  \includegraphics[width=\columnwidth]{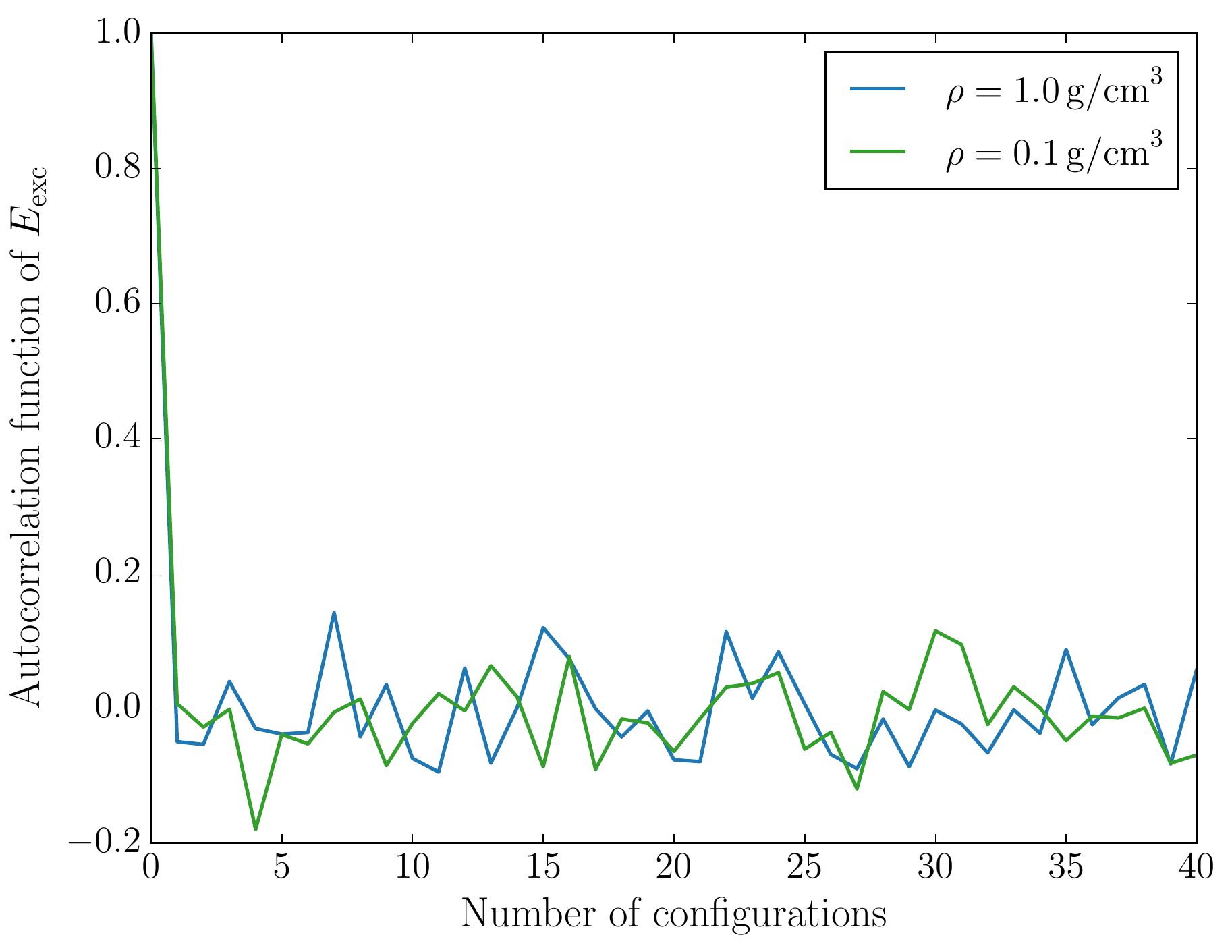}
  \caption{Autocorrelation function of the excess energy time series shown in Figure 
    \ref{fig:eexc_vs_nsnap}.}
  \label{fig:autocorr_eexc}
\end{figure}

\begin{figure}
  \centering
  \includegraphics[width=\columnwidth]{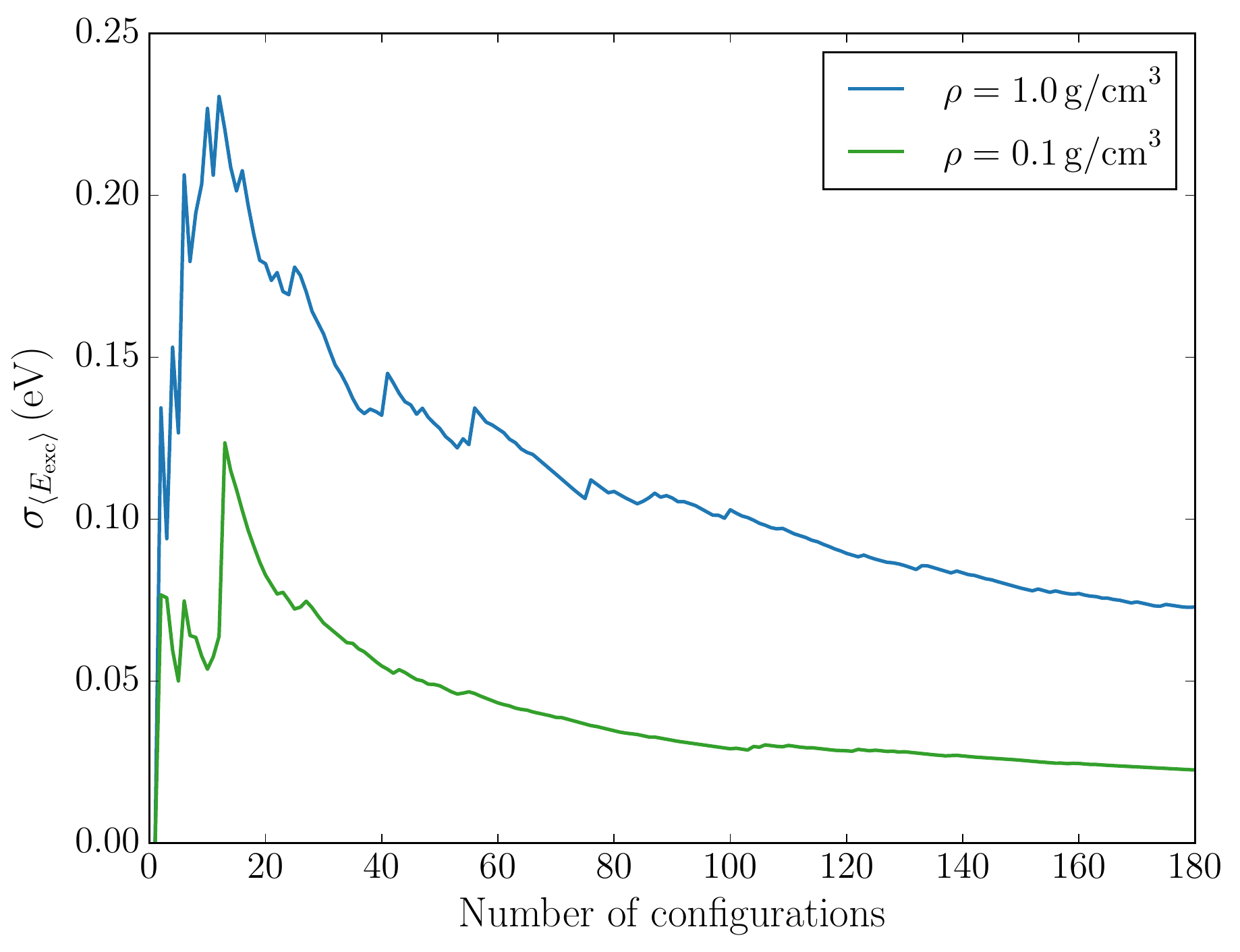}
  \caption{Standard error of the mean of the Ca excess energy at $T=4000\,$K
    with respect to the number of independent configurations used to compute the mean, 
    for different helium densities.}
  \label{fig:stderror_eexc}
\end{figure}

When computing the excess energy $E_{\mathrm{exc}}$ using
configuration snapshots extracted from MD simulations, the results can
fluctuate drastically from one configuration to the other. This is shown in
Figure \ref{fig:eexc_vs_nsnap}, where the lines represent the
evolution of $E_{\mathrm{exc}}$ from configuration to
configuration. In Figure \ref{fig:autocorr_eexc}, we show the
autocorrelation function of the $E_{\mathrm{exc}}$ time series,
\begin{equation}
  r_k = \frac{\sum_{i=1}^{N-k} \left( E_{\mathrm{exc}}^i - 
    \langle E_{\mathrm{exc}} \rangle \right) \left( E_{\mathrm{exc}}^{i+k} - 
    \langle E_{\mathrm{exc}} \rangle \right)}{\sum_{i=1}^{N} \left( E_{\mathrm{exc}}^i - 
    \langle E_{\mathrm{exc}} \rangle \right)^2}.
\end{equation}
Since the autocorrelation function quickly decays to zero, we conclude
that the time elapsed between each configuration snapshot is long
enough for the $E_{\mathrm{exc}}$ time series values to be
statistically independent. Therefore, we can safely apply the
central-limit theorem to compute the standard error of the mean,
\begin{equation}
  \sigma_{\langle E_{\mathrm{exc}} \rangle} = \frac{\sigma_{E_{\mathrm{exc}}}}{\sqrt{N}}.
\end{equation}
Figure \ref{fig:stderror_eexc} shows the evolution of $\sigma_{\langle
  E_{\mathrm{exc}} \rangle}$ with respect to the number of
configurations used to compute the mean. For both $\rho=0.1$ and
$\rho=1.0\,{\rm g\,cm}^{-3}$, we notice the $1/\sqrt{N}$ decay of
$\sigma_{\langle E_{\mathrm{exc}} \rangle}$. This implies that to
improve the error by a factor of two, the number of configurations
needs to be quadrupled. From this analysis, we chose to use 100
configurations for each $(T,\rho)$ condition. This value is enough to
obtain $\sigma_{\langle E_{\mathrm{exc}} \rangle} \lesssim 0.1\,$eV
for most physical conditions considered in this work, which is an
error that we consider acceptable for our purpose.

\paragraph{Validation with ab initio molecular dynamics}
\label{sec:abinitio_md}
Since our $\phi_{\mathrm{Ca-He}}(r)$ potential was calculated in the
infinite-dilution limit, one could be worried about the exactitude of
the atomic configurations obtained through molecular dynamics using
this potential. To check this point, we computed the excess
energy of Ca using configurations extracted from \textit{ab initio} molecular
dynamics simulations.  In this framework, no pair potential is
assumed. The electronic density, energy and forces on ions are
recomputed at every time step of the simulation using DFT. This
approach is expected to be more exact than the classical molecular
dynamic approach, but its computational cost is larger by orders of
magnitude. These calculations were performed using Born-Oppenheimer
molecular dynamics with the CPMD
package\footnote{\url{http://cpmd.org}} \citep{cpmd,marx2000ab},
with the PBE exchange-correlation functional and ultrasoft
pseudopotentials \citep{vanderbilt1990soft}. We employed 0.5\,fs time
steps and an energy cut-off of 35\,Ry. As before, we extracted atomic
configurations from these simulations and used these configurations to
compute the interaction energy of Ca with the surrounding medium
through DFT calculations.

\begin{figure}
  \centering
  \includegraphics[width=\columnwidth]{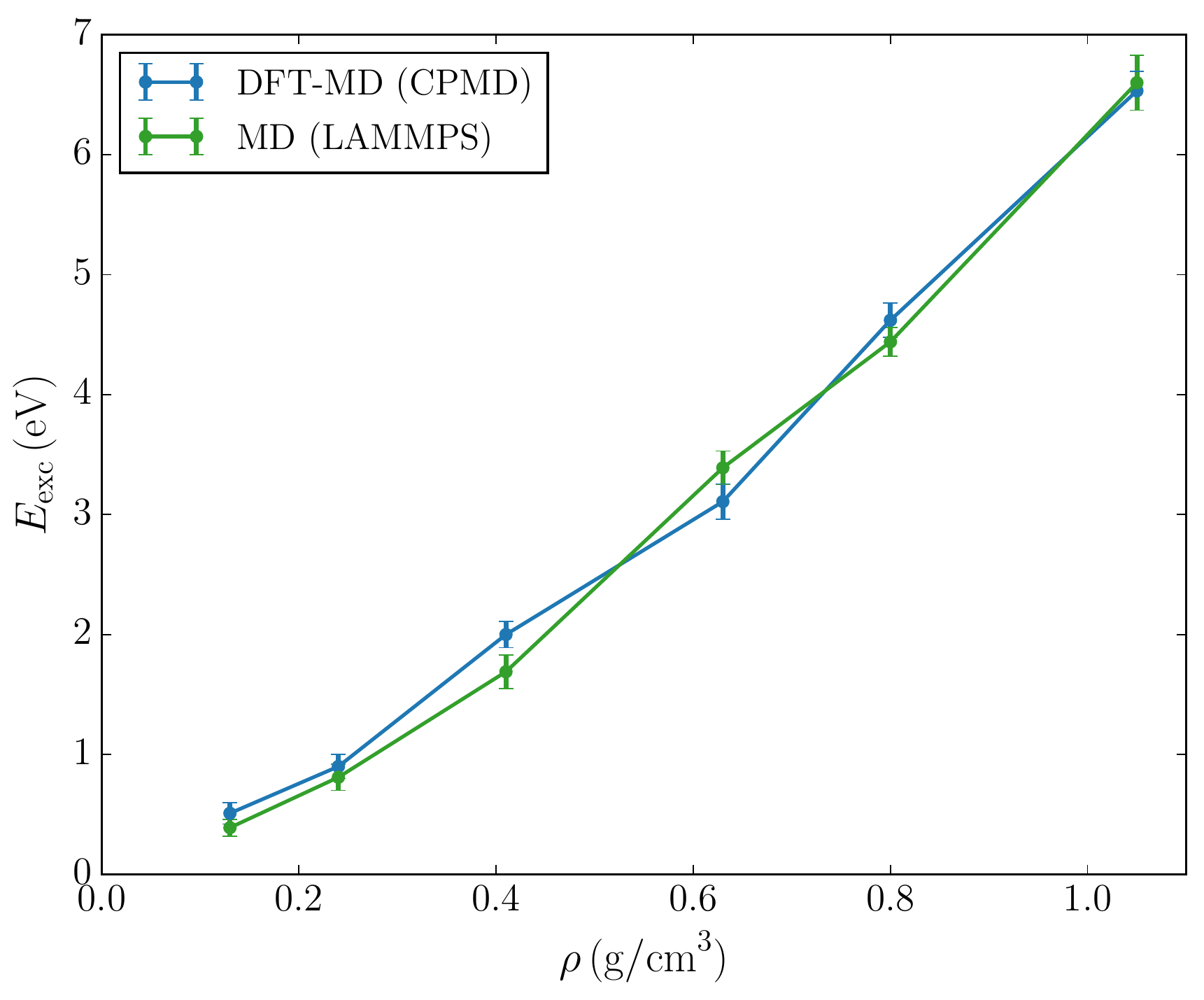}
  \caption{Excess energy of Ca at $T=5000\,$K for different helium densities, obtained from 
    configurations extracted from ab initio molecular dynamics (DFT-MD) and from 
    classical molecular dynamics (MD) using the pair potentials described in 
    Section \ref{sec:excluded_volume}.}
  \label{fig:cpmdVSlammps}
\end{figure}

Figure \ref{fig:cpmdVSlammps} compares the results obtained to those
found with the classical molecular dynamics simulations. This
comparison shows that there is only a negligible difference between
the two approaches, at least below $\rho=1\,{\rm g\,cm}^{-3}$. We did
not perform any comparison at higher densities, because of the
prohibitive calculation time of such calculations. In any case,
densities above $1\,{\rm g\,cm}^{-3}$ are never encountered at the
photosphere of cool DZ white dwarfs (Section
\ref{sec:models}). Therefore, we conclude that our infinite-dilution
limit potential $\phi_{\mathrm{Ca-He}}(r)$ is sufficient to generate
the atomic configurations used to compute the excess energy (and
it is much faster than resorting to ab initio molecular dynamics
simulations).

\subsubsection{Ionization equilibrium} 
\label{sec:final_results}
\begin{figure}
  \begin{center}
    \includegraphics[width=\columnwidth]{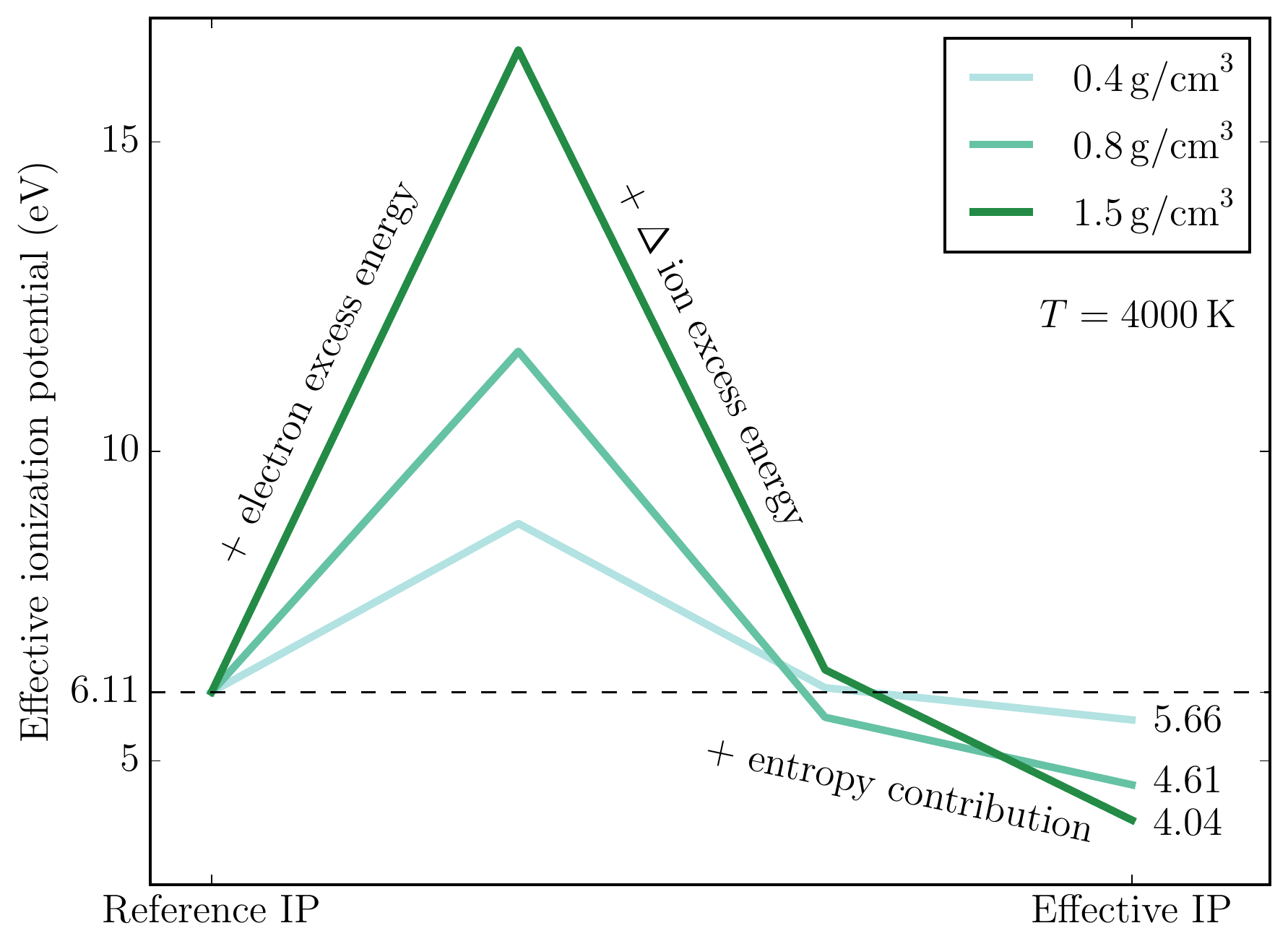}
  \end{center}
  \caption{Contributions added to the reference ionization potential of Ca to obtain its 
    effective ionization potential at various densities (see legend). 
    These results are for \mbox{$T=4000\,{\rm K}$}.}
\label{fig:plot_evol}
\end{figure}

Following the methodology described in the previous sections, we
computed $\mu_K^{\mathrm{nid,ent}}$ and $E_K^{\mathrm{exc}}$ for
\ion{C}{1}/\ion{C}{2}, \ion{Ca}{1}/\ion{Ca}{2},
\ion{Fe}{1}/\ion{Fe}{2}, \ion{Mg}{1}/\ion{Mg}{2} and
\ion{Na}{1}/\ion{Na}{2}. By adding these excess chemical potentials to
the electron excess energy, we computed by how much the ionization
potential is altered at a given density and temperature (Equation
\ref{eq:deltaI}). Figure \ref{fig:plot_evol}, which shows the three
contributions to $\Delta I$ (the free electron excess energy, the
variation of $E_K^{\mathrm{exc}}$ and the change in
$\mu_K^{\mathrm{nid,ent}}$), illustrates this process in the case of
Ca.

\begin{figure*}
  \begin{center}
    \includegraphics[width=\linewidth]{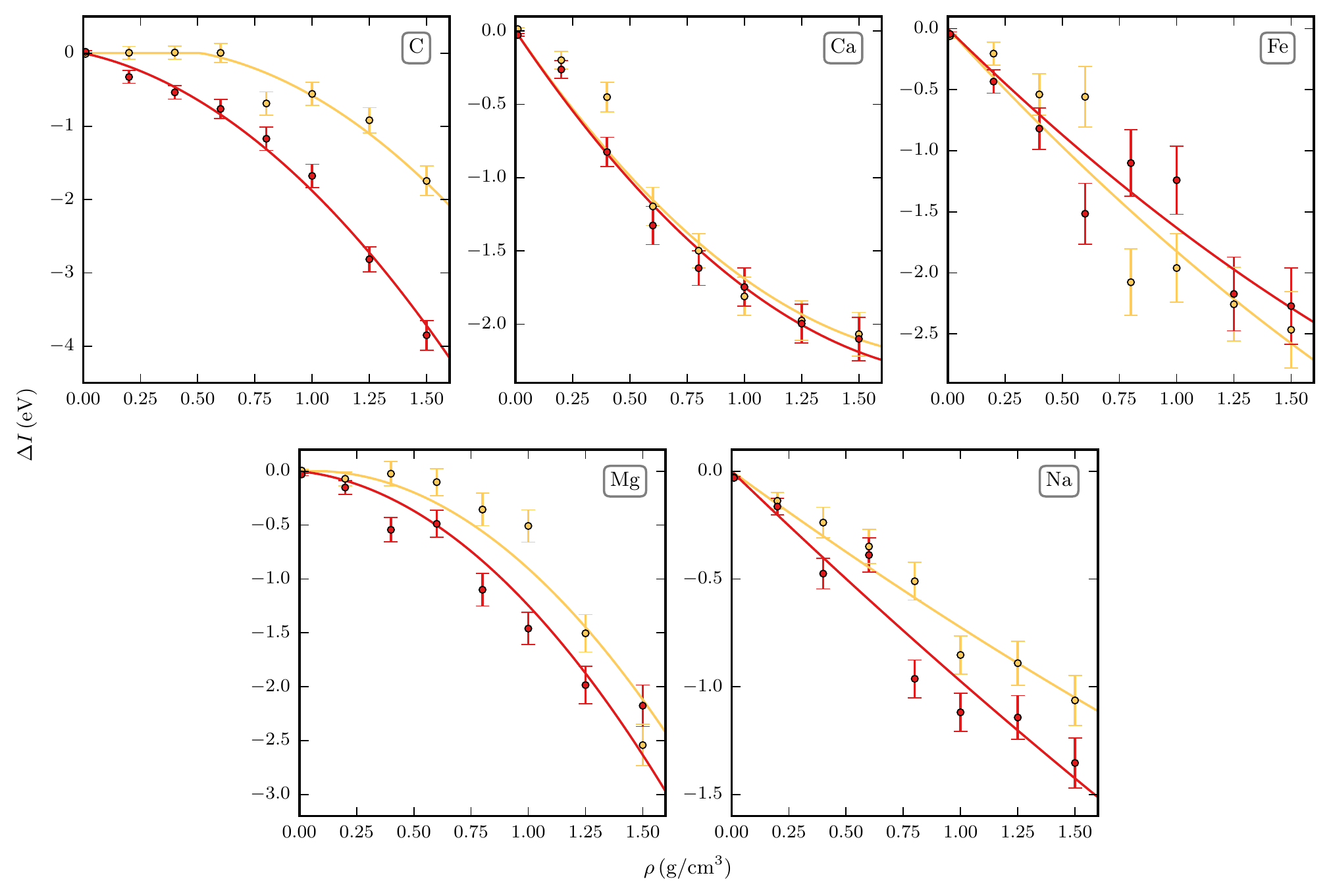}
  \end{center}
  \caption{Depression of the ionization potential of C, Ca, Fe, Mg and Na embedded in a dense 
    helium fluid. Circles show the results of our ab initio calculations and error bars indicate
    the statistical errors associated with the configuration sampling. The solid lines show 
    the analytical fits found through a chi-squared minimization of Equation 
    \ref{eq:fitting_function}. Data in red are for $T=8000\,$K and data in yellow are for 
    $T=4000\,$K.}
  \label{fig:mega_delta}
\end{figure*}

Figure \ref{fig:mega_delta} shows our final results. First, for every
ion considered, we notice that $\Delta I \rightarrow 0$ when $\rho
\rightarrow 0$. This is the expected behavior and it shows that our
methodology is consistent with the ideal regime when we push it to low
densities.  Secondly, we note that $\Delta I$ is always negative and
that its absolute value increases with density. This result means that
ionization becomes easier with increasing density, which also
corresponds to the expected behavior. Finally, for all elements except
Fe, we notice that higher temperatures are associated with slightly
larger ionization potential depressions. This result is consistent
with the findings of \cite{kowalski2007equation}, who found a
reduction of the band gap of warm dense helium with increasing
temperature.

To easily implement these nonideal ionization potentials in atmosphere
models, we have fitted our results with a simple function of $\rho$
and $T$,
\begin{equation}
  \Delta I(\rho,T) = \min \left \{ 0, (a + b T) \rho + c \rho^2 \right \},
  \label{eq:fitting_function}
\end{equation}
where $a$, $b$ and $c$ are parameters found using a chi-squared
minimization algorithm, $\rho$ is the helium density in ${\rm
  g\,cm}^{-3}$ and $T$ is the temperature in K. This expression allows
a satisfactory fit to the data and yields $\Delta I =0$ at
$\rho=0$. The analytical fits are shown in Figure \ref{fig:mega_delta}
and the fitting parameters are reported in Table \ref{tab:fitparams}.
Formally, in order to stay within the limits of our calculations, the use 
of these analytical expressions should be limited to densities between 0 
and $1.5\,{\rm g\,cm}^{-3}$ and to temperatures between 4000 and 8000$\,{\rm K}$.
Nevertheless, we have verified that Equation \ref{eq:fitting_function} 
can safely be extrapolated to lower (down to 2000$\,{\rm K}$) or higher 
temperatures (at least up to 10000$\,{\rm K}$) if needed.

\begin{deluxetable}{crrr}
  \tablecaption{Fitting parameters for $\Delta I (\rho,T)$ (Equation \ref{eq:fitting_function}). 
    \label{tab:fitparams}}
  \tablehead{\colhead{Ion} & \colhead{$a^1$} & \colhead{$b^2$} & \colhead{$c^3$} }
  \startdata
  C  &  1.91782 & -3.24813 & -1.19948 \\
  Ca & -2.20703 & -0.14431 &  0.57494  \\
  Fe & -2.23142 &  0.48427 &  0.21301  \\
  Mg &  0.45809 & -0.85522 & -1.01958  \\
  Na & -0.52305 & -0.62471 &  0.04833 
  \enddata
  \tablenotetext{1}{${\rm eV\,g}^{-1}{\rm \,cm}^3$}
  \tablenotetext{2}{$10^{-4}\,{\rm eV\,g}^{-1}{\rm \,K}^{-1}{\rm \,cm}^3$}
  \tablenotetext{3}{${\rm eV\,g}^{-2}{\rm \,cm}^6$}
\end{deluxetable}

\subsubsection{Comparison with previous studies}

It is instructive to compare these results with the ionization
equilibrium predicted by the Hummer-Mihalas occupation probability
formalism, which is widely used in atmosphere codes.  Since there is
no theoretical prescription for the values of the hard sphere radii
used to compute the occupation probabilities (Equation \ref{eq:hm}), a
somewhat arbitrary choice must be made to perform this comparison. We
chose to compute the hard sphere radii with the hydrogenic
approximation, as described by \cite{beauchampPhD}. In this
approximation, the radius of a species in state $i$ is given by
\begin{equation}
  r_i = \frac{n_i^2 a_0}{Z^{\rm eff}_i},
\end{equation}
where $n_i$ is the principal quantum number of the uppermost electron, $a_0$ is the Bohr radius 
and the effective nuclei charge $Z^{\rm eff}_i$ is given by
\begin{equation}
  Z^{\rm eff}_i = n_i \sqrt{\frac{I_i}{13.598\,{\rm eV}}},
\end{equation}
where $I_i$ is the energy needed to ionize an electron from state
$i$. In the Hummer-Mihalas formalism, every term in the partition
function is multiplied by the occupation probability (Equation
\ref{eq:partHM}). If we stick to the ground-state approximation
(Section \ref{sec:approximations}), the occupation probability is the
same for every level and it can be factored out of the partition
function sum. Hence, the net effect of the Hummer-Mihalas formalism is
to multiply the right-hand side of the Saha equation (Equation
\ref{eq:saha_ideal}) by the ratio of occupation probabilities, $w_{\rm
  Z II}/w_{\rm Z I}$.

\begin{figure}
  \includegraphics[width=\columnwidth]{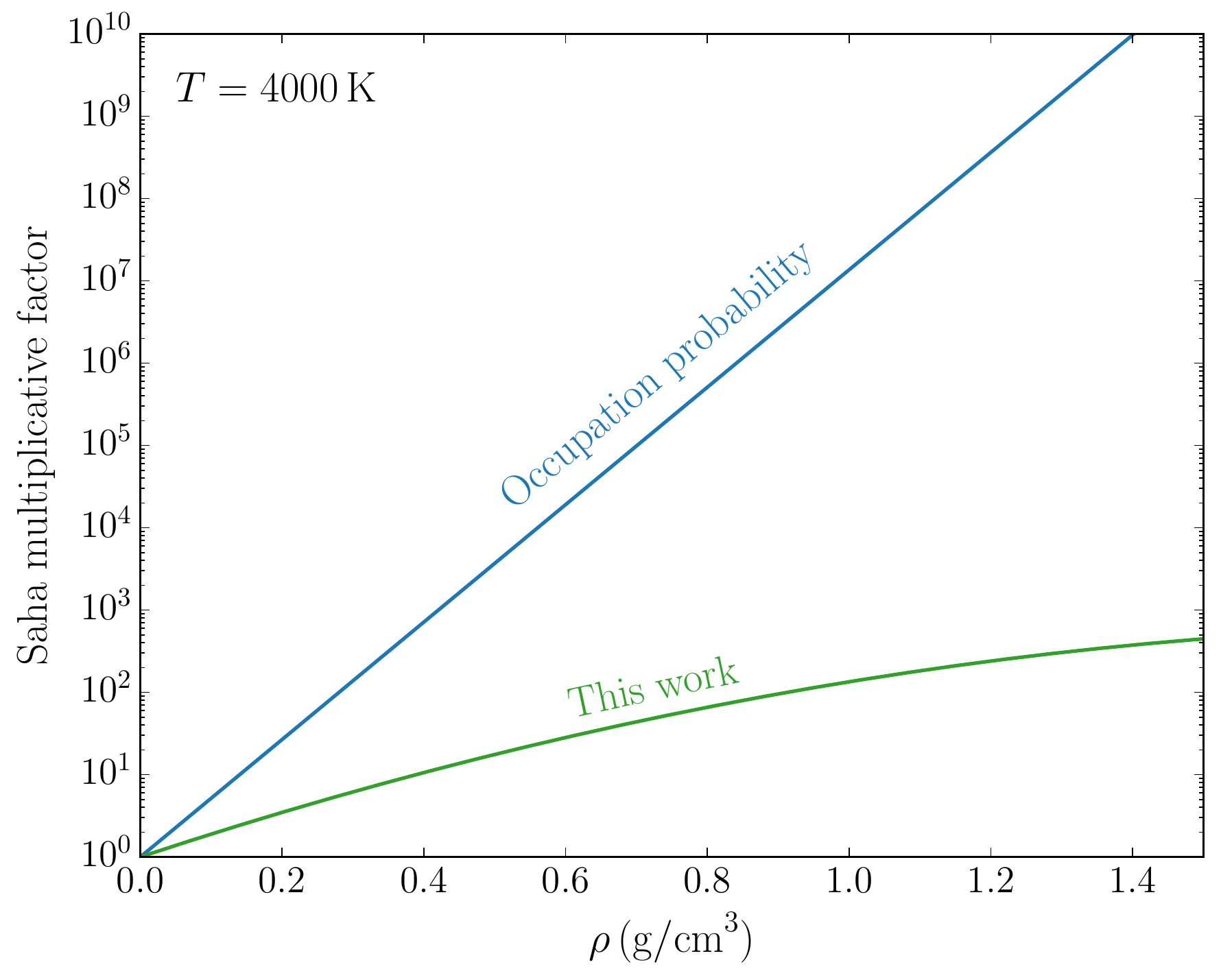}
  \caption{Multiplicative factor applied to the right-hand side of
    the \ion{Ca}{1}/\ion{Ca}{2} Saha equation (Equation \ref{eq:saha_ideal})
    to take nonideal effects into account. The blue line is 
    $w_{\rm Ca II}/w_{\rm Ca I}$, the result obtained using the Hummer-Mihalas formalism, 
    and the green curve is $e^{-\Delta I/(k_B T)}$, the result obtained with our ionization
    model.}
  \label{fig:comp_hm}
\end{figure}

Figure \ref{fig:comp_hm} compares the multiplicative factors that need
to be applied to the right-hand side of the Saha equation for the
\ion{Ca}{1}/\ion{Ca}{2} ionization equilibrium to account for nonideal
effects (i.e., $w_{\rm Ca II}/w_{\rm Ca I}$ in the case of the
Hummer-Mihalas formalism and $e^{-\Delta I/(k_B T)}$ for our
ionization model). The most obvious aspect of Figure \ref{fig:comp_hm}
is that we find a weaker pressure ionization than what is predicted
using the Hummer-Mihalas formalism and hard sphere radii computed in
the hydrogenic approximation. We checked that this result holds true
for C, Fe, Mg and Na.  This conclusion is consistent with the findings
of \cite{bergeron1991synthetic} for the ionization equilibrium of
hydrogen in cool DA stars. Using the Hummer-Mihalas formalism and a
hydrogen radius given by $r_n = n^2 a_0$, they found that the high
Balmer lines are predicted to be too weak, indicating that pressure
ionization in the Hummer-Mihalas formalism is too strong. They showed
that using a smaller radius in the computation of the occupation
probabilities, $r_n = 0.5 n^2 a_0$, allows better spectral fits.

Unfortunately, we cannot compute the ionization potential depression
of H to directly confirm the conclusion of
\cite{bergeron1991synthetic}. The problem is that the \ion{H}{2}--He
potential \citep[e.g.,][]{pachucki2012born,kolos1976new} has a deep
attractive well (since H$^+$ and He can form the HeH$^+$ molecule)
that prevents proper convergence of the OZ equation solver. The same
issue arises if we try to compute the ionization potential of H in a
H-rich medium, since the \ion{H}{2}--\ion{H}{1} potential
\citep[e.g.,][]{frost1954semiempirical} also has an important
attractive well (H$^+$ and H can form the H$_2 ^+$ molecule).

\subsection{Atmosphere models}
\label{sec:models}

\begin{figure}
  \includegraphics[width=\columnwidth]{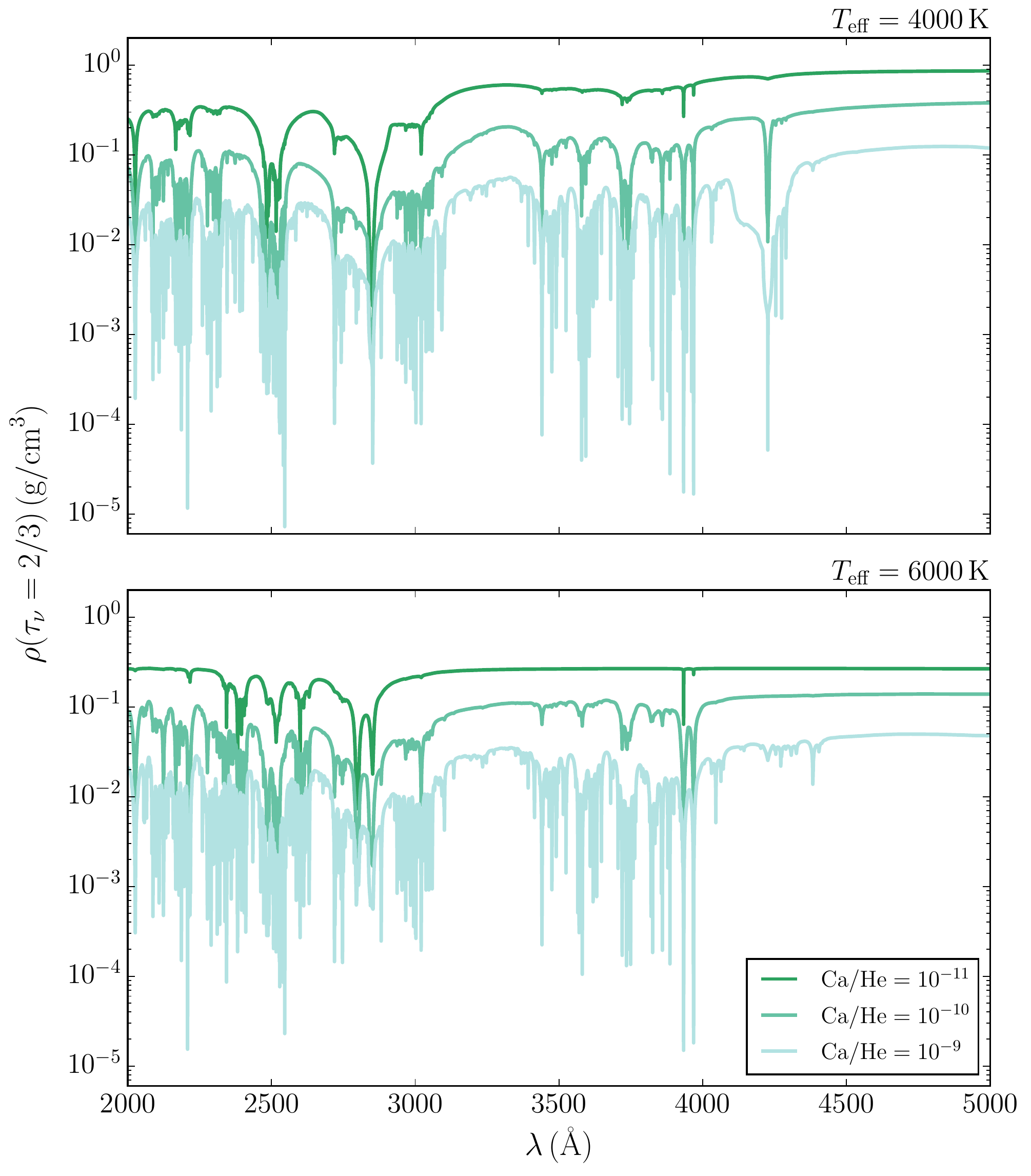}
  \caption{Density at an optical depth $\tau_{\nu}=2/3$ with respect to $\lambda$. The top 
    panel shows the results for $T_{\rm eff}=4000\,{\rm K}$ models and the bottom panel for 
    $T_{\rm eff}=6000\,{\rm K}$. The Ca abundance is given in the legend and a surface 
    gravity $\log g=8$ is assumed.}
  \label{fig:rhotau}
\end{figure}

Using the analytical model described in the previous Section, we
implemented the improved ionization equilibrium of heavy elements in
our atmosphere code to investigate how it affects the synthetic
spectra of cool DZ stars.  Before even examining any spectrum, we can
get an idea of the impact of the new nonideal ionization equilibrium
by looking at the densities involved in the model atmospheres.  Figure
\ref{fig:rhotau} shows the density at $\tau_{\nu}=2/3$ as a function
of $\lambda$ for a few atmosphere models with different effective
temperatures and calcium abundances.
\footnote{In this paper, the abundance of all metallic species, from C
  to Cu, is scaled to the abundance of Ca to match the abundance
  ratios of chondrites reported in \cite{lodders2003solar}.}  This
type of figure is useful to identify which densities are probed at
different wavelengths. In the previous Section, we saw that no
important deviation from the ideal ionization equilibrium is expected
below $0.1\,{\rm g\,cm}^{-3}$ (see Figure \ref{fig:mega_delta}). From
Figure \ref{fig:rhotau}, it is clear that the probed densities are
below this threshold for ${\rm Ca/He} \gtrsim 10^{-10}$ and above this
threshold for ${\rm Ca/He} \lesssim 10^{-10}$. Therefore, it should
become important to take into account the nonideal ionization
equilibrium for cool DZ atmosphere models with ${\rm Ca/He} \lesssim
10^{-10}$, but it is probably superfluous for models with ${\rm Ca/He}
\gtrsim 10^{-10}$ (note that nonideal effects on the opacities
and the equation of state remain nevertheless important in this regime).
For intermediate densities (${\rm Ca/He} \approx
10^{-10}$), using the nonideal ionization equilibrium should result in
small changes in the spectral line wings of the coolest models.

\begin{figure*}
  \includegraphics[width=\textwidth]{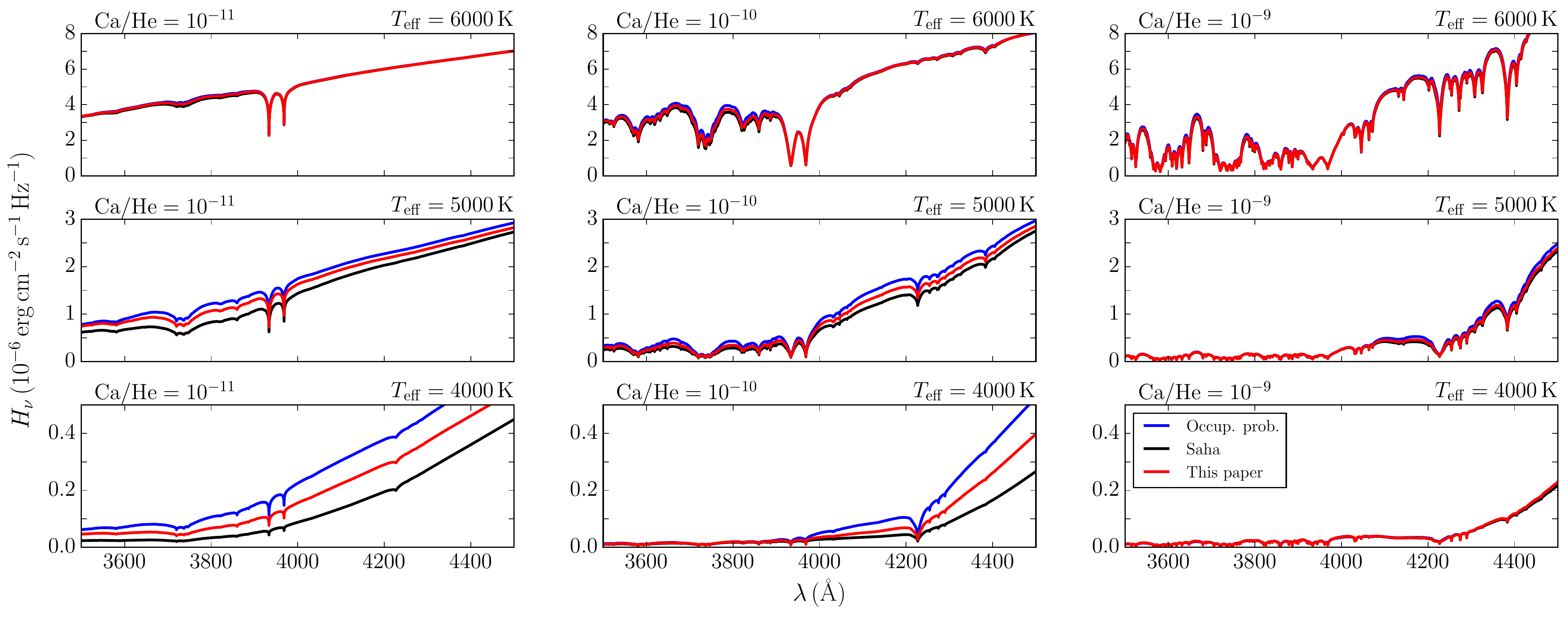}
  \caption{Comparison between synthetic spectra computed using the \cite{hummer1988equation} 
    formalism (in blue), the 
    ionization equilibrium presented in this work (in red) and the ideal Saha equation (in black). 
    All models 
    were computed assuming $\log g=8$ and ${\rm H/He}=0$. The effective temperature and 
    the metal abundance is indicated above each panel.}
  \label{fig:comp_spec_ioniz}
\end{figure*}

Figure \ref{fig:comp_spec_ioniz} compares synthetic spectra computed
with our ionization equilibrium model to spectra computed using the
occupation probability formalism and the ideal Saha equilibrium (in
each case, the atmosphere model structure and the synthetic spectrum
were computed using the same ionization model).  This figure focuses
on the region between $3500$ and $4500\,$\AA, since it contains
several Ca, Fe and Mg absorption lines susceptible of being affected
by the choice of the ionization model.  The first thing to note is that for the
high-density models (i.e., those with a low metal abundance and a low
effective temperature) there are important differences between spectra
obtained using the ideal Saha equilibrium and our ionization model.
These differences are mostly due to a shift in the continuum
associated with the increased electronic density in models that take
pressure ionization into account.  Next, we notice that the spectra
computed using the Hummer-Mihalas formalism are even further from the
spectra obtained using the ideal Saha equilibrium than those computed
with our ionization model.  This is not surprising, since as seen in
Figure \ref{fig:comp_hm}, the Hummer-Mihalas formalism predicts a very strong
pressure ionization.  Finally, for the low-density models (i.e., those
with a high metal abundance and/or a high effective temperature), all
three sets of spectra are virtually identical, which is consistent
with our analysis of Figure \ref{fig:rhotau}.

The nonideal chemical equilibrium of heavy elements also has a small impact on
the model atmosphere structure. The increased electronic density
associated with pressure ionization leads to an increase of the
Rosseland mean opacity and therefore to a reduction of the pressure at
the photosphere.  For instance, for $T_{\rm eff}=4000\,{\rm K}$, $\log
g = 8$ and ${\rm Ca/He} = 10^{-11}$, a model that assumes the ideal
Saha equation has a photospheric density of $0.93\,{\rm g\,cm}^{-3}$,
while an atmosphere structure based on our ionization model has a
photospheric density of $0.89\,{\rm g\,cm}^{-3}$. Moreover, the
occupation probability formalism predicts a density that is still
lower ($0.84\,{\rm g\,cm}^{-3}$). Given Figure \ref{fig:comp_hm}, this
result is not surprising: compared to our calculations, the
Hummer-Mihalas formalism overestimates the efficiency of pressure
ionization.

Our results constitute a physically-grounded answer to the question of
the importance of pressure ionization in cool DZ stars, which will
help to reduce the gap between solutions found with different
atmosphere codes. A good example to illustrate this point is vMa2 (WD
0046+051).  On one hand, using an ideal treatment of chemical
equilibrium, \cite{dufour2007spectral} found a solution with $T_{\rm
  eff}=(6220 \pm 240) \,{\rm K}$. On the other hand, using the Hummer-Mihalas
occupation probability formalism, \cite{wolff2002element} found
$T_{\rm eff}=(5700 \pm 200) \,{\rm K}$.  In their analysis,
\cite{dufour2007spectral} showed that the difference between both
solutions can largely be explained by the different chemical
equilibrium models used in both studies.  This uncertainty can be
removed by relying on the accurate description of the chemical
equilibrium described in the current work.

\section{Applications}
\label{sec:applications}

To show how the improved constitutive physics presented in this work
translates in terms of better spectroscopic fits, this
Section presents the analysis of two well-known DZ stars: Ross 640
(WD 1626+368) and LP 658-2 (WD 0552-041). Applications to other objects
will be presented in other papers of the series.

Our new analysis of these two objects makes use of \textit{Gaia} DR2
parallaxes \citep{prusti2016gaia,brown2018gaia}, $BVRI$ and $JHK$
photometry published in (\citealt{bergeron2001photometric}, see Table
\ref{tab:obs_data}), optical spectra published in
\cite{giammichele2012know} and UV spectra obtained with HST and the
Faint Object Spectrograph (FOS, \citealt{koester2000element,wolff2002element})

\begin{deluxetable}{cccc}
  \tablecaption{Observational data. \label{tab:obs_data}}
  \tablehead{\colhead{} & \colhead{Ross 640} & \colhead{LP 658-2}}
  \startdata
  Parallax (mas)  &  $62.915 \pm 0.022$ & $155.250 \pm 0.029$ \\
  $B^{1}$   & $14.02 $ & $15.49 $ \\ 
  $V$       & $13.83 $ & $14.45 $ \\
  $R$       & $13.75 $ & $13.99 $ \\
  $I$       & $13.66 $ & $13.54 $ \\
  $J$       & $13.58 $ & $13.05 $ \\
  $H$       & $13.57 $ & $12.86 $ \\
  $K$       & $13.58 $ & $12.78 $ 
  \enddata
  \tablenotetext{1}{There is a 3\% uncertainty on all photometric measurements.}
\end{deluxetable}

\subsection{Ross 640}
\label{sec:ross640}

At $T_{\rm eff} \approx 8000\,{\rm K}$, Ross 640 is technically not a
"cool" white dwarf. Since the density at its photosphere is $\approx
0.01\,{\rm g\,cm}^{-3}$ ($n_{\rm He} = 1.5 \times 10^{21}\,{\rm
  cm}^{-3}$), nonideal effects affecting the equation of state and the
chemical equilibrium are minimal. However, this density is high enough
to induce important differences between Lorentzian profiles and the
improved line profiles presented in Section
\ref{sec:line_profiles}. This object is therefore the perfect
candidate to test our line profiles separately, without the
interference of other nonideal effects.

To fit this star, we follow the procedure described in \cite{dufour2007spectral}.
In short, we first find $T_{\rm eff}$ and $\log g$ using
the photometric technique described in
\cite{bergeron2001photometric}. The photometric measurements are
first converted into fluxes using the constants reported in
\cite{holberg2006calibration}. Then, these observed fluxes $f_{\nu}$
are compared to the model fluxes $H_{\nu}$ to obtain $T_{\rm eff}$
and the solid angle $\pi (R/D)^2$, where $R$ is the radius of the star
and $D$ is its distance to the Earth.  These parameters are found
using a $\chi^2$ minimization technique relying on the
Levenberg-Marquardt algorithm. Since $D$ is known from the parallax
measurement, the radius $R$ can be computed from the solid
angle. The mass of the star and the corresponding surface gravity $g=GM/R^2$
are then found using the evolutionary models of \cite{fontaine2001potential}.
This $\log g$ value being generally different from our initial
guess, we repeat the fitting procedure until all fitting parameters
are converged.

Once a consistent solution for $T_{\rm eff}$ and $\log g$ is obtained
from the procedure described in the previous paragraph, we move to
the determination of the abundances using spectroscopic
observations. We keep $T_{\rm eff}$ and $\log g$ fixed to the values
found using the photometric observations and then fit the Ca/He and
H/He ratios by minimizing the $\chi^2$ between our synthetic spectra
and the observed spectrum. Since the abundances found with this
technique are generally different from those initially used for the
photometric fit, the whole fitting procedure is repeated until
internal consistency is reached.

\begin{figure}
  \includegraphics[width=\columnwidth]{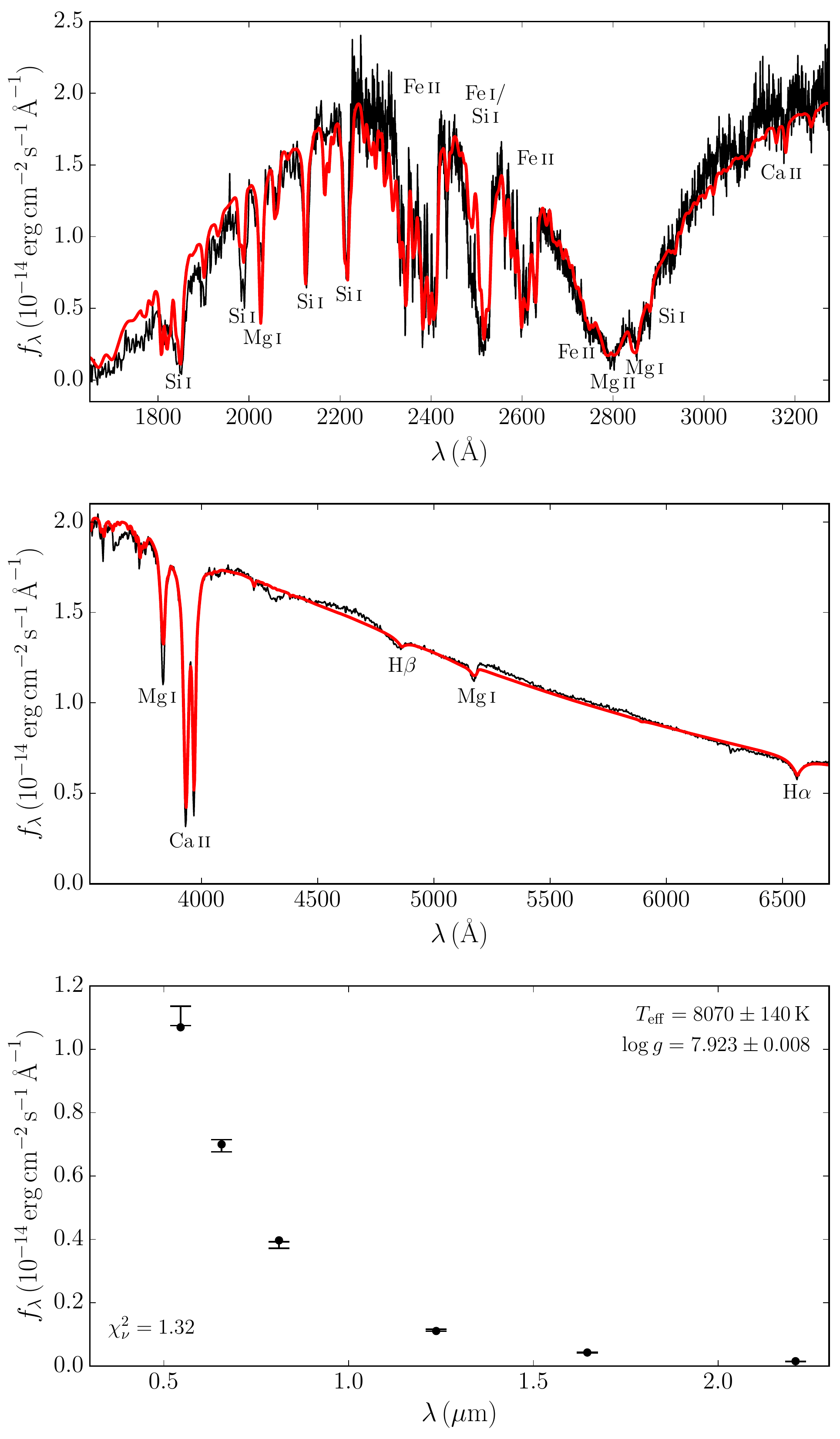}
  \caption{Our best solution for Ross 640. The top panel shows our fit 
    to the UV spectrum, the middle panel is our fit to the visible spectrum 
    and the bottom panel shows our photometric fit to the $BVRI$ and 
    $JHK$ bands.}
  \label{fig:ross640}
\end{figure}

Although the abundance ratio between the different heavy elements is
kept constant during the $\chi^2$ minimization procedure, we manually
adjust the abundance ratio of Mg, Fe and Si to fit the spectral
lines labeled in Figure \ref{fig:ross640}.  All other heavy elements
(from C to Cu) are included in the models, but since we could not use
any spectral line to fit their abundances, we simply assume the same
abundance ratio with respect to Ca as in chondrites
\citep{lodders2003solar}.

As shown in Figure \ref{fig:ross640}, our solution is consistent with
observations across all wavelengths. Our fitting parameters, given in
Table \ref{tab:fit_sol}, are roughly similar to those found by
\cite{dufour2007spectral}, \cite{koester2000element} and \cite{zeidler1986metal}
, although they all found a higher effective temperature ($8440 \pm
320\,{\rm K}$, $8500 \pm 200\,{\rm K}$ and $8800\,{\rm K}$, respectively).  
One major improvement
compared to previous authors is our fit to the broad \ion{Mg}{2}
2795/2802\,{\AA} lines. To obtain a good fit,
\cite{koester2000element} arbitrarily multiplied the van der Waals
broadening constant of these lines by 10. No arbitrary constants are
needed using our new line profiles and a consistent abundance is found
from both the optical and ultraviolet magnesium lines.

\begin{deluxetable}{ccc}
  \tablecaption{Fitting parameters.
    \label{tab:fit_sol}}
  \tablehead{\colhead{} & \colhead{Ross 640} & \colhead{LP 658-2}}
  \startdata
  $T_{\rm eff}\,{\rm (K)}$ & $\phm{-} \phd 8070 \pm 140 \phd \phn$    & $\, \phd 4430 \pm 40 \phn \,$   \\
  $\log g$                 & $\phm{-} 7.923 \pm 0.008$ & $\phm{-} 7.967 \pm 0.022$ \\
  $\log {\rm H/He}$        & $-3.5 \pm 0.2$    & $<-5$   \\
  $\log {\rm Ca/He}$       & $-9.12 \pm 0.05$    & $-11.38 \pm 0.05\;\;$\\
  $\log {\rm Fe/He}$       & $-8.44 \pm 0.10$    & -   \\
  $\log {\rm Mg/He}$       & $-7.40 \pm 0.10$    & $-8.66 \pm 0.20$ \\
  $\log {\rm Si/He}$       & $-7.90 \pm 0.20$    & -  
  \enddata
\end{deluxetable}

\subsection{LP 658-2}
\label{sec:lp658-2}

LP 658-2 is a DZ star that exhibits a weak \ion{Ca}{2} H \& K doublet.
During the last two decades, many authors have tried to fit this star, but none has reached
a consistent solution across all wavelengths. Because they relied on different models and 
observations, the solutions they found are quite diverse (see Table \ref{tab:lp658-2}).

\begin{deluxetable}{ccc}
  \tablecaption{Literature review of LP 658-2. \label{tab:lp658-2}}
  \tablehead{\colhead{Authors} & \colhead{$T_{{\rm eff}}$ (K)} & \colhead{H/He}}
  \startdata
  \cite{bergeron2001photometric} & $5060 \pm 60$ & He \\
  \cite{wolff2002element}        & $5060 \pm 60$ & ${\rm H/He} = 5 \times 10^{-4}$\\
  \cite{dufour2007spectral}      & $4270 \pm 70$ & He \\
  \cite{giammichele2012know}     & $5180 \pm 80$ & H 
  \enddata
\end{deluxetable}

First, \cite{bergeron2001photometric} found that LP 658-2 has a
helium-rich atmosphere with $T_{\rm eff}=(5060 \pm 60)\,{\rm
  K}$. However, their analysis was based on atmosphere models that did
not include heavy elements, which strongly influence UV opacities and the
temperature profile.

Then, using HST data (FOS), \cite{wolff2002element} extended the
analysis of \cite{bergeron2001photometric} with an investigation of
the UV portion of the spectrum of LP 658-2.  The large absorption
feature observed in the UV was interpreted as strong broadening from
the wing of Ly$\alpha$. Keeping the effective temperature fixed
at the $T_{\rm eff}=5060\,{\rm K}$ value found by
\cite{bergeron2001photometric}, they used this UV absorption feature
to fit the hydrogen abundance and found that ${\rm H/He} = 5 \times
10^{-4}$. However, contrarily to other stars in their sample (e.g.,
LHS 1126 and BPM 4729), they were not able to properly reproduce the
shape of this UV absorption feature.

Subsequently, using models that include heavy elements in the
atmosphere structure, \cite{dufour2007spectral} determined a much
cooler temperature for LP 658-2 ($T_{\rm eff}=4270 \pm 70 \,{\rm
  K}$). At this temperature, the photometric data can completely
exclude the presence of traces of hydrogen at the level found by
\cite{wolff2002element} since H$_2$-He CIA would cause a strong IR
flux depletion that is not observed. However, the solution of
\cite{dufour2007spectral} does not explain the UV absorption
feature seen in the FOS data and their spectroscopic solution
predicted a large \ion{Ca}{1} 4226\,{\AA} line, which is completely absent
from the observations.

More recently, \cite{giammichele2012know} argued that the narrow H \& K
lines observed in the spectra of LP 658-2 indicate that it is perhaps
a hydrogen-rich star after all. However, although an H-rich
composition allowed a better fit to the visible spectrum than
\cite{dufour2007spectral}, the photometric fit was not as good (and
it can not explain the shape of the UV spectrum).

\begin{figure}
  \includegraphics[width=\columnwidth]{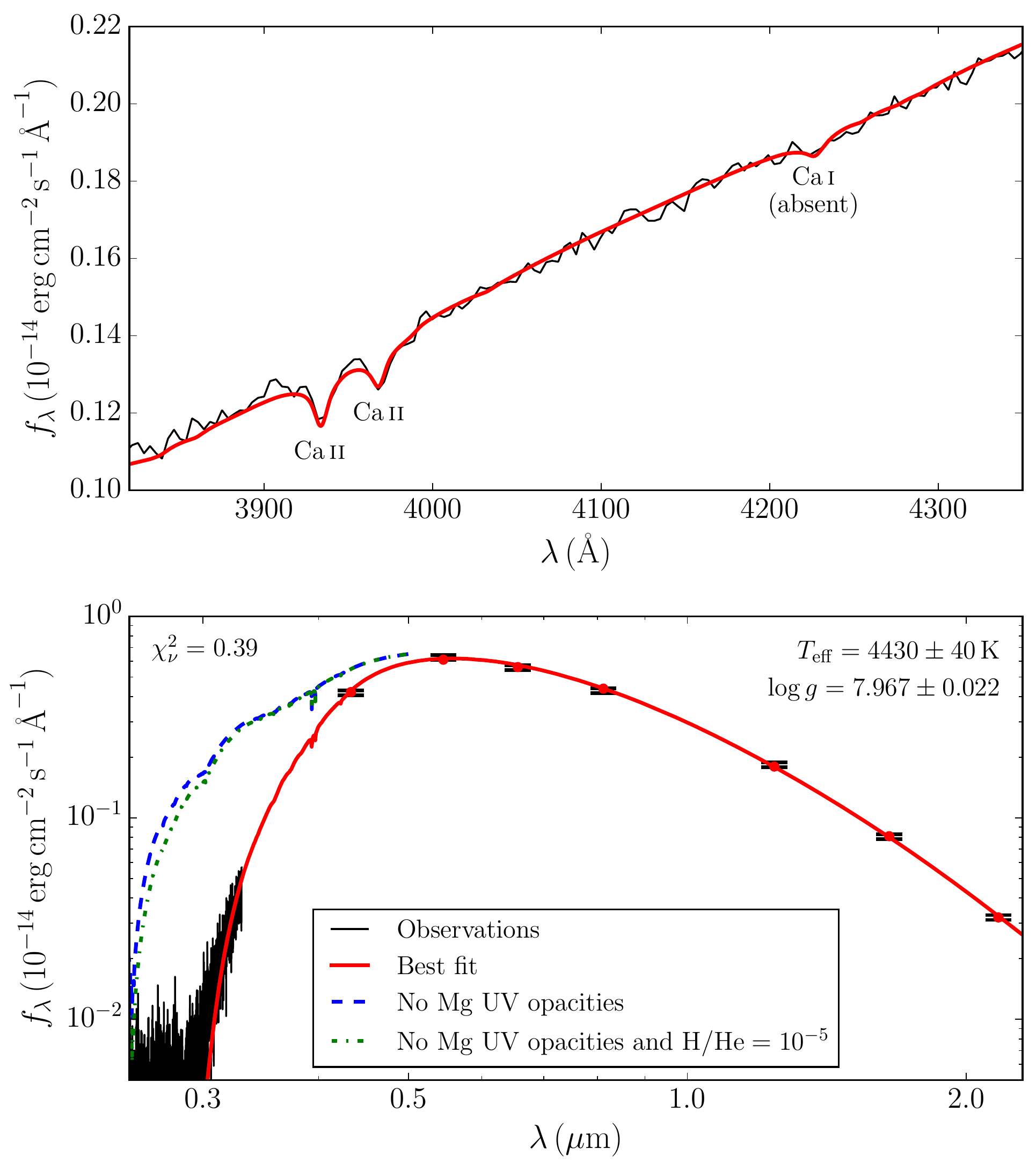}
  \caption{Our best solution for LP 658-2. The top panel shows our fit to 
    the visible spectrum and the bottom panel displays our fit to the 
    photometric observations and to the FOS data. The bottom panel also shows 
    two synthetic UV spectra computed without the \ion{Mg}{2} 2795/2802\,{\AA} 
    and the \ion{Mg}{1} 2852\,{\AA} lines, one without hydrogen (in blue) and 
    one with ${\rm H/He}=10^{-5}$ (in green).}
  \label{fig:lp658-2}
\end{figure}

Using our improved models, we can now obtain a solution that agrees perfectly
with the observations across all wavelengths assuming a helium-rich
atmosphere (Figure \ref{fig:lp658-2}). We can also constrain the
amount of hydrogen to ${\rm H/He} < 10^{-5}$, as a higher hydrogen
abundance would produce an IR flux depletion that is incompatible with
the observations. Given this limit, the shape of the UV continuum can
no longer be explained by the wing of Ly$\alpha$ (see the green dash-dot
line in Figure \ref{fig:lp658-2}). Instead, we find
that the absorption in the UV can naturally be explained by the
presence of trace amounts of magnesium (absorption from the
\ion{Mg}{2} 2795/2802\,{\AA} and the \ion{Mg}{1} 2852\,{\AA}
lines). While there is formally no lines detected, the amount of
magnesium needed to reproduce the UV continuum is small enough as to
not produce features in the optical spectrum.

Finally, our new models do not predict the strong \ion{Ca}{1}
4226\,{\AA} line that was predicted using the models of
\cite{dufour2007spectral}. This is mainly due to the use of our
improved line profiles (Section \ref{sec:line_profiles}) as well as
our new nonideal Ca ionization equilibrium calculation (Section
\ref{sec:ionization_main}), the former effect being the most
important. Our fitting parameters, given in Table
\ref{tab:fit_sol}, were found using the same fitting procedure as for
Ross 640.

\section{Conclusion}
\label{sec:conclusion}

We have developed an updated atmosphere model code that incorporates all the 
necessary constitutive physics for an accurate description of cool DZ stars. 
This code includes
\begin{itemize}
\item The most important heavy element line profiles computed using
  the unified line shape theory of \cite{allard1999effect},
\item CIA profiles suitable for fluids where the density exceeds $0.1\,{\rm g\,cm}^{-3}$,
\item He Rayleigh scattering and He$^-$ free-free absorption corrected
  for collective interactions between atoms,
\item An ab initio equation of state for H and He,
\item A nonideal chemical equilibrium model for He, C, Ca, Fe, Mg and Na.
\end{itemize}
While most of these nonideal effects were implemented using results previously published
by various authors, we performed our own calculations to assess the chemical
equilibrium of heavy elements. 

More precisely, we used the classical theory of fluid and DFT
calculations to characterize the ionization equilibrium of C, Ca, Fe,
Mg and Na in a dense helium medium and under the temperature and density
conditions found in the atmosphere of cool DZ stars. These
calculations show that the effective ionization potential begins to
decrease when the density exceeds $0.1\,{\rm g\,cm}^{-3}$, reaching a
depression of $\approx 1-2\,{\rm eV}$ at $\rho = 1\,{\rm
  g\,cm}^{-3}$. We provided analytical fits to our data that can be
implemented in atmosphere model codes to obtain the effective
ionization potential for a given temperature and density.

We computed atmosphere models using this improved description of the
ionization of heavy elements and found that under the right conditions (i.e.,
weakly polluted, low-$T_{\rm eff}$ objects) the synthetic spectrum can
significantly differ from results obtained using the ideal Saha
equation.  Moreover, we found that the Hummer-Mihalas formalism --
when used in conjunction with hydrogenic hard sphere radii -- leads to
a much stronger pressure ionization than our model, which indicates an
overestimation of pressure ionization. This result is consistent with
previous findings based on comparisons between atmosphere models and
observed spectra \citep{bergeron1991synthetic}.  Finally, we showed
how the improved constitutive physics included in our code
translates into better spectral fits for Ross 640 and LP 658-2, two
cool DZ stars that presented a challenge to previous atmosphere model
codes.

In the next papers of this series, we will use our updated models to analyze
in detail other cool white dwarfs, in particular WD 2356-209 (a
peculiar cool DZ star showing an exceptionally strong Na D feature) and the
first cool DZ star to show CIA absorption. We will also analyze the bulk of
the known cool white dwarfs taking advantage of the \textit{Gaia} data and
revisit the spectral evolution of these objects.

\acknowledgments We wish to thank Piotr M. Kowalski for useful
discussions regarding the DFT calculations presented in Section
\ref{sec:ionization_main}. This work was supported in part by NSERC
(Canada).

This work has made use of data from the European Space Agency (ESA) mission
{\it Gaia} (\url{https://www.cosmos.esa.int/gaia}), processed by the {\it Gaia}
Data Processing and Analysis Consortium (DPAC,
\url{https://www.cosmos.esa.int/web/gaia/dpac/consortium}). Funding for the DPAC
has been provided by national institutions, in particular the institutions
participating in the {\it Gaia} Multilateral Agreement.

This work has made use of the Montreal White Dwarf Database \citep{dufour2016montreal}.

This work used observations made with the NASA/ESA Hubble Space
Telescope, and obtained from the Hubble Legacy Archive, which is a
collaboration between the Space Telescope Science Institute
(STScI/NASA), the Space Telescope European Coordinating Facility
(ST-ECF/ESA) and the Canadian Astronomy Data Centre (CADC/NRC/CSA).

%\software{CPMD \citep{marx2000ab}, LAMMPS \citep{plimpton1995fast}, 
%Matplotlib \citep{hunter2007matplotlib}, NumPy \citep{walt2011numpy}, 
%ORCA \citep{neese2012orca}, {\sc Quantum ESPRESSO} \citep{giannozzi2009quantum}}

\bibliographystyle{aasjournal}
\bibliography{references}

\end{document}